\documentclass[aps,twocolumn,groupedaddress,amsmath,amssymb]{revtex4}
\usepackage{graphicx}  
\usepackage{dcolumn}   
\usepackage{bm}        
\usepackage{verbatim}   
\usepackage{graphicx,amsmath}
\usepackage{subfigure}
\usepackage{mathrsfs}
\usepackage{graphics}
\usepackage{epstopdf}
\usepackage{hyperref}
\usepackage{amsmath}
\usepackage{amssymb}
\usepackage{dcolumn}
\usepackage{slashed}

\begin{document}

\title{Quasinormal modes of scalar and Dirac perturbations of
Bardeen de-Sitter black holes}
\author{Wadbor Wahlang\footnote{Email: wadbor@iitg.ernet.in}}
\author{Piyush A. Jeena\footnote{Email: p.jeena@iitg.ernet.in}}
\author{S. Chakrabarti\footnote{Email: sayan.chakrabarti@iitg.ernet.in}}

\affiliation{ Department of Physics,\\
   Indian Institute of Technology Guwahati\\
Guwahati-781039, India}

\begin{abstract}
So far the study of black hole perturbations has been mostly focussed upon the classical black holes with singularities at the origin and hidden by  event horizon. Compared to that, the regular black holes are a completely new class of solutions arising out of modification of general theory of relativity by coupling gravity to an external form of matter. Therefore it is extremely important to study the behaviour of such regular black holes under different types of perturbations. Recently a new regular Bardeen black hole solution with a de Sitter branch has been proposed by Fernando \cite{fern2}. We compute the quasi-normal (QN) frequencies for the regular Bardeen de Sitter (BdS) black hole due to massless and massive scalar field perturbations as well as the massless Dirac perturbations. We analyze the behaviour of both real and imaginary parts of quasinormal frequencies by varying different parameters of the theory. 

\end{abstract}

\maketitle

\section{Introduction}
In the study of physics, the stability criteria of a system or configuration is one of the main interesting aspects. 
Unstable system or configurations are generally not realizable in nature and they are generally an
intermediate stage in the dynamical evolution of a system. A black hole system in general relativity can also 
be put in the above mentioned category: the question one asks there is whether a black hole which is stable
under some perturbation, i.e. if we perturb the black hole from outside, whether it comes back to its original 
state after some time or whether the perturbation grows unbound making the black hole unstable.\par

The study of black hole perturbations remains an extremely 
intriguing topic which has enormous effect on various important properties 
of a black hole \cite{kk,nol,v1,kon1}. In general, the dynamical evolution of perturbations of a black hole background 
can be classified into three stages, the first of which consists of an initial outburst of wave,
depending completely on the initial perturbing field,
the second stage consists of damped oscillations, known in the literature as the quasinormal modes (QNM) whose 
frequency turns out to be a complex number, the real part representing the oscillation frequency and 
the imaginary part representing damping. QN frequencies completely depend on the background and on the 
nature of the perturbation and thereby giving immense importance to these modes which are used
to determine the black hole parameters (mass, charge and angular momentum). Thus the QN frequencies are encoding the information about the relaxation of a black hole which has been perturbed. The third is a power law tail behaviour at very late times. The 
equations governing the black hole perturbations in most of the cases can be cast into a Schr\"{o}dinger like wave equation. The QNMs are 
solutions to the wave equation with complex frequencies with a boundary condition which are completely ingoing at the horizon and purely
outgoing at asymptotic infinity. In the present work we will be focussing on the second of the above three stages of evolution of black hole perturbation in a regular black hole background in asymptotically de Sitter space-time. 


The importance of studying black holes in de Sitter space lies in the fact that our universe looks like asymptotically de Sitter at very early and 
late times. Recent observational data also indicates that our universe is going through a phase of accelerated expansion \cite{perl,riess,tegmark_sdss}, thereby providing the existence of a positive cosmological constant. In general de Sitter space turns out to be a maximally symmetric solution to the vacuum Einstein equations with a positive cosmological constant. The perturbations and stability of black holes  in de Sitter space have been studied extensively and there have been a lot of work \cite{kono3}-\cite{sm1} on quasinormal modes of scalar, electromagnetic, gravitational and Dirac perturbations, decay of charged fields, asymptotic quasinormal modes and signature of quantum gravity. 

In the study of perturbations of black holes, the examples so far mainly focus on the black hole geometries with singularities at the origin. In general theory of relativity, the existence of singularity is an inherent feature of all physically relevant classical black hole solutions. Penrose's cosmic censorship conjecture ensures that the singularity must be hidden by an event horizon, thereby preventing the pathologies occurring at the singularity to influence the exterior region of the black hole. However, it is expected that a modification of general theory of relativity (be it quantum or classical) may be able to rectify the pathological behaviour of the classical black hole solutions.  When a black hole does not have a space-time singularity at the origin, it is termed as a ``regular black hole'' in the literature. Since we do not have a complete theory of quantum gravity, regular black hole solutions may be constructed by coupling gravity with external matter fields. The first solution of such regular black holes  with non-singular geometry satisfying the weak energy condition was obtained by Bardeen \cite{bardeen}, which is now known as the Bardeen black hole. However, the solution Bardeen proposed lacked physical motivation  because the solution was not a vacuum solution, rather gravity was modified by introducing some form of matter and thereby introducing an energy momentum tensor in the Einstein's equation. The introduction of the energy-momentum tensor was done in an {\it ad hoc} manner. Much later, Ay\'{o}n-Beato and Garc\'{i}a \cite{abg1} showed the energy momentum tensor to be the gravitational field of some magnetic monopole arising out of a specific form of non-linear electrodynamics. Subsequently, many other solutions \cite{abg2}-\cite{bronn2}, motivating the avoidance of singularity was proposed in the literature. Therefore, although the regular black holes presently do not have any observational signature and are presently treated as toy models to see how to avoid pathologies due to singularity at the interior of a black hole, it is extremely important as well as relevant to study how the QN frequencies behave for such regular black holes and see how differently they respond to the perturbations, compared to the usual classical black holes with singularities at the interior. There were many works published regarding such regular black holes: stability properties \cite{ms1}, QNMs \cite{nino,fern1}, thermodynamics \cite{man} and geodesic structure \cite{zcw} of regular black holes to mention a few. Very recently Fernando \cite{fern2} has proposed a  de Sitter branch for the regular Bardeen black hole and calculated the grey body factor for such a black hole. In this paper, we will be  discussing the QNMs of the Bardeen de Sitter (henceforth BdS) black hole due to scalar (both massless and massive) and Dirac perturbations. Although study of scalar field perturbations in a black hole background and its corresponding QNMs is not new, the Dirac field perturbations, on the other hand, are relatively less studied. Therefore, apart from the scalar perturbations,  it will also be interesting to study the Dirac perturbations in the regular black hole backgrounds in de Sitter space.\par

The plan of the paper is as follows: in the next section we briefly discuss the BdS black hole. In section-III, we present a brief discussion of WKB method along with a study of the scalar QNMs of the BdS black holes. Section-IV deals with the Dirac quasinormal modes of the BdS black hole. Finally, in section-V we conclude the paper with a brief discussion on future directions.

\section{A discussion on BdS black hole}
In this section we will briefly discuss about the Bardeen de Sitter (BdS) black hole following the works of Fernando \cite{fern2}. 
The author of this paper modified the works of \cite{abg1} to incorporate a positive cosmological constant in the action:
\begin{equation}
S=\int d^4x\sqrt{-\tilde g}\left(\frac{R-2\Lambda}{16\pi}-\frac{1}{4\pi}\mathcal{L}(F)\right),\label{action}
\end{equation}
where $\tilde g$ is the determinant of the metric tensor, $R$ is the Ricci Scalar and $\mathcal{L}(F)$ is the function of the field strength tensor of the non-linear electrodynamics $F_{\mu\nu}=2(\nabla_{\mu} A_{\nu}-\nabla_{\nu}A_{\mu})$ and its form is  given by 
\begin{equation}
\mathcal{L}(F)=\frac{3}{2\alpha g^2}\left(\frac{\sqrt{2g^2F}}{1+\sqrt{2g^2F}}\right)^{5/2}.
\end{equation}
In the above, the parameter $\alpha$ is related to the magnetic charge and the mass of the black hole in the following manner: $\alpha=\frac{|g|}{2M}$.  As was mentioned in the introduction, the Bardeen black hole as proposed initially, was not an exact solution to Einstein equations and hence there was no known sources of physical origin associated with it. The quantity $g$ was left as a regularizing parameter of the theory without any physical interpretation being associated with it. Later on Ay\'{o}n-Beato and Garc\'{i}a \cite{abg1} had provided the regular Bardeen black hole model with a physical interpretation. It was shown by them \cite{abg1} that the regularizing parameter $g$ can be physically interpreted as the monopole charge of a self-gravitating magnetic field of nonlinear electrodynamics. 

If one derives the equations of motion from the above action(\ref{action}), then following equations will be arrived at
\begin{align}
G_{\mu\nu}+\Lambda {\tilde g}_{\mu\nu}&=2\left(\frac{\partial \mathcal{L}(F)}{\partial F}F_{\mu\lambda}F^{\lambda}_{\nu}-{\tilde g}_{\mu\nu}\mathcal{L}(F)\right)& \\
\nabla_{\mu}\left(\frac{\partial \mathcal{L}(F)}{\partial F}F^{\nu\mu}\right)&=0\\
\nabla_{\mu}(* F^{\nu\mu})&=0
\end{align}
It was shown in \cite{fern2} that a static spherically symmetric solution for the above set of equations exist:
\begin{equation}
ds^2=-f(r)dt^2+f(r)^{-1}dr^2+r^2(d\theta^2+sin^2\theta d\phi^2)\label{metric}
\end{equation}
with $f(r)$ being given 
\begin{equation}
f(r)=1-\frac{2Mr^2}{(r^2+g^2)^{3/2}}-\frac{\Lambda r^2}{3}. \label{fr}
\end{equation}
From the metric function one can find the asymptotic behaviour as $f(r)\sim 1-2M/r+3g^2 M/r^3 +\mathcal{O}(1/r^5)-\Lambda r^2/3$.
The $1/r$ term dictates that the parameter $M$ must be associated with the mass of the configuration, which can also be
verified from the explicit evaluation of the ADM mass. However, the next term goes as $1/r^3$ and hence, unlike the Reissner-Nordstr\"{o}m case, this does not allow one to associate the parameter $g$ with the `Coulomb' charge. The importance of the term was realized later by Ay\'{o}n-Beato and Garc\'{i}a \cite{abg1} and subsequently a physical interpretation of $g$ was motivated as discussed above. 
The solution of $f(r)=0$ gives the horizon and in the particular case of BdS black hole there may be  three real roots implying three horizons: the black hole inner and outer horizons along with the cosmological horizon. There lies the possibility of getting either one real root corresponding to cosmological horizon only for a set of parameters of this theory or a possibility of getting degenerate roots corresponding to a merger of the inner and outer black hole horizons for a range of parameters $M, g$ and $\Lambda$. Structurally the BdS black hole is similar to the Reissner-Nordstr\"{o}m-de Sitter (RNdS) or Born-Infeld de Sitter (BIdS) black holes which also admit a possibility of three distinct horizons as well as a single or degenerate horizons. However, it was shown in \cite{fern2} that the event horizon is larger in the case of RNdS black hole compared to a BdS one. 
The interesting nature of BdS geometry is its non-singular structure everywhere. It can be checked by direct calculation that all the scalar curvatures $R$, $R_{\mu\nu}R^{\mu\nu}$, $R_{\mu\nu\lambda\sigma}R^{\mu\nu\lambda\sigma}$ are finite everywhere except for the electromagnetic field invariant $F$ which is singular at $r=0$ \cite{fern2}.  

\section{QNMs of massless and massive scalar perturbations in BdS black hole}
In this section we will consider the massless and massive scalar field perturbations of the BdS black hole geometry to study 
the behaviour of the QNMs in BdS background with the given black hole parameters. 
  \subsection{Massless scalar field perturbation}
  In general, when one discusses the perturbation of a black hole from a  theoretical view point, there are  two different ways to initiate the perturbation - one is by adding test fields to the black hole geometry and the other is by perturbing the black hole metric itself, since in general relativity spacetime does not merely act as a stage where dynamics happen, rather spacetime itself is a dynamical quantity. When a test field does not backreact (i.e. in the linear approximation) on the background geometry, the first kind of perturbation can be reduced to the study of propagation of fields in the black hole background, which is a general covariant equation of motion of the corresponding field. The covariant form of the equation of motion is different for different spin fields in a curved background. The simplest of the ways to study black hole perturbation due to external fields is to study the scalar (spin $s=0$) wave equation in a black hole background. The reason being its simplicity as well as the fact that the tensor type gravitational perturbations always coincides with the massless uncharged scalar field perturbations \cite{v1}. In case of the BdS black holes also, we will follow the same route to have a quick glance at the nature of the QN frequencies. 
  
  As discussed in Section-II,  BdS background metric is given by equations (\ref{metric}) and (\ref{fr}).
 The Klein-Gordon equation for  a massless scalar field $\Phi$ is
   \begin{equation}
   \nabla^2\Phi=0
  \end{equation}
which explicitly takes the form
 \begin{equation}
  \frac{1}{\sqrt{-\tilde g}} \partial_{\mu}\left(\sqrt{-\tilde g}\tilde g^{\mu\nu}\partial_\nu \Phi\right)=0 
 \end{equation}
 As usual, we introduce the ansatz for $\Phi$ as,
 \begin{equation}
  \Phi=e^{-i\omega t}Y_{\ell,m}(\theta,\phi) \frac{U(r)}{r}.\label{ansatz}
 \end{equation}
 With the above ansatz,we have the standard Schr\"{o}dinger-like wave equation for the 
 perturbation of the BdS metric by a scalar field is given by
\begin{equation}
 \frac{d^2U(r)}{dr_*^2}+\left(\omega^2-V_{\rm{scalar}}(r_*)\right)U(r)=0
\end{equation}
where, \hspace{0.2cm}$V_{\rm{scalar}}(r)=f(r)\left(\frac{\ell(\ell+1)}{r^2}+\frac{f'(r)}{r}\right)$. The coordinate $r_*$ is the standard tortoise coordinate related to radial coordinate $r$ 
as  $dr_*=\frac{dr}{f(r)}$. The advantage of using the tortoise coordinate lies in the fact 
that the range of the coordinate now extends between $-\infty$ to $\infty$, whereas in the old radial
coordinate $r$, the physically accessible region lies between the black hole and cosmological horizon.
Note also that the potential $V_{\rm{scalar}}(r)\to 0$ as $r_{*}\to \pm\infty$.  
\begin{figure}[!hb]
\centering
\includegraphics[ scale=0.65]{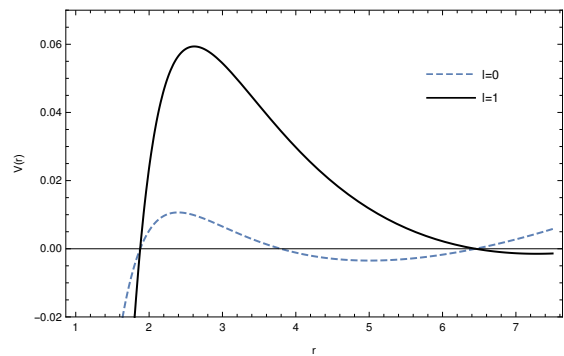}
\caption{Variation of effective potential $V_{\rm{scalar}}(r)$ vs $r$ for $\ell=0, 1$ with $M=1$ , $g=0.55$ , $\Lambda=0.05$. Note the negative minimum between the cosmological and black hole horizon in the $\ell=0$ plot. }
\label{fig1}
\end{figure}
It can be easily seen by plotting the scalar field potential against the radial coordinate for various values
of the multipole number $\ell$ that the $\ell=0$ mode has a distinct local minimum between the black hole outer
horizon and the cosmological horizon (see \figurename{\ref{fig1}}), which was also pointed out in \cite{fern2}. For this reason, the method used in this paper to evaluate the QNMs for the BdS black hole, namely the WKB approach is not a valid one to evaluate QNMs for $\ell=0$ modes. Therefore, from now on, we will only talk about $\ell\neq 0$ modes for the massless scalar QNMs of BdS black hole.

As already stated, we will solve the wave equation for complex QN frequencies semi-analytically, using the sixth order WKB method developed in \cite{kon5}. It has been shown extensively in literature that WKB method works extremely well for determining QN frequencies. The sixth order WKB method is more accurate than the third order method and the former in fact gives results practically coinciding with those obtained from full numerical integration of the wave equation \cite{kon5} for low overtones, i.e. for modes with small imaginary parts, and for all multipole numbers $\ell\ge 1$. The sixth order formula for a general black hole potential $V(r)$ is mentioned below
\begin{equation}
 \frac{i(\omega^2-V(r_0))}{\sqrt{-2V^{''}(r_0)}}-\Lambda_2-\Lambda_3-\Lambda_4-\Lambda_5-\Lambda_6 = n+\frac{1}{2}\label{qnmeqn}
\end{equation}
where $V(r_0)$ is peak value of $V(r)$ , $V^{''}(r_0) = \frac{d^2V}{dr_*^2}|_{r=r_0}$ , $r_0$ is the value 
of the radial coordinate corresponding to the maximum of the potential $V(r)$ and $n$ is the overtone number.
QN frequencies $\omega$ would be of the form $\omega=\omega_R+\omega_I$. In eqn.[\ref{qnmeqn}], $\Lambda_2$ and $\Lambda_3$ are given by \cite{will2}
\begin{align}
 \Lambda_2=&\frac{1}{\sqrt{2V^{''}(r_0)}}\Big[\frac{1}{8}\left(\frac{V_0^{(4)}}{V^{''}(r_0)}\right)(b^2+\frac{1}{4})\nonumber\\
 &-\frac{1}{288}\left(\frac{V_0^{(3)}}{V^{''}(r_0)}\right)^2(7+60b^2)\Big]\\
 \Lambda_3=&\frac{(n+\frac{1}{2})}{2V^{''}(r_0)}\Big[\frac{5}{6912} \left( \frac{V_0^{(3)}}{V^{''}(r_0)} \right)^4(77+188b^2)\nonumber\\
 &-\frac{1}{384} \left( \frac{(V_0^{(3)})^2 V_0^{(4)}}{(V^{''}(r_0))^3}\right) (51+100b^2)\nonumber\\
 &+ \frac{1}{2304}\left( \frac{V_0^{(4)}}{V^{''}(r_0)}\right)^2(67+68b^2)\nonumber\\
 &+ \frac{1}{288}\left(\frac{V_0^{(3)}V_0^{(5)}}{(V^{''}(r_0))^2}\right)(19+28b^2)\nonumber\\
 &-\frac{1}{288} \left(\frac{V_0^{(6)}}{V^{''}(r_0)}\right)(5+4b^2)\Big].
\end{align}

\begin{figure}[htbp]
\centering
  \centering
  \includegraphics[scale=0.8]{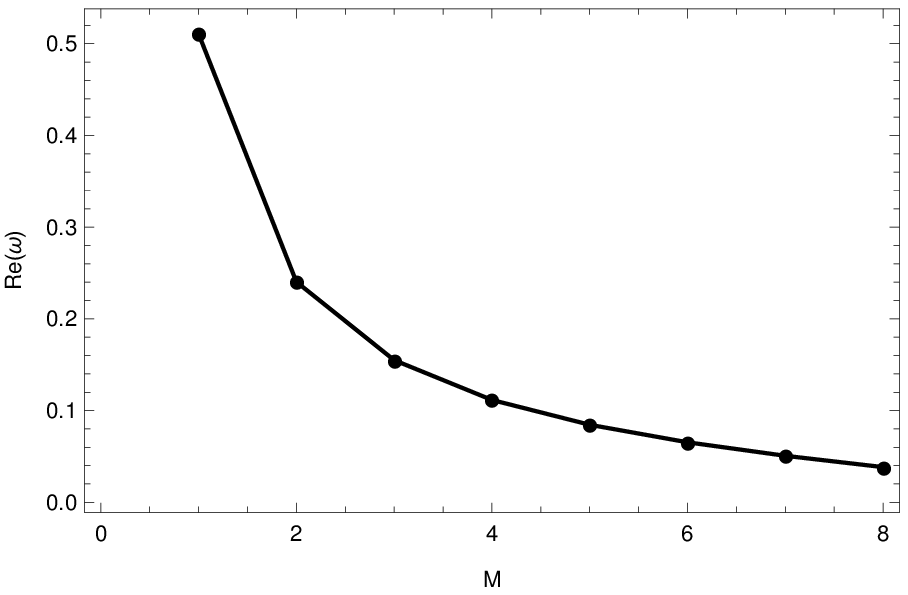}\\
  \includegraphics[scale=0.8]{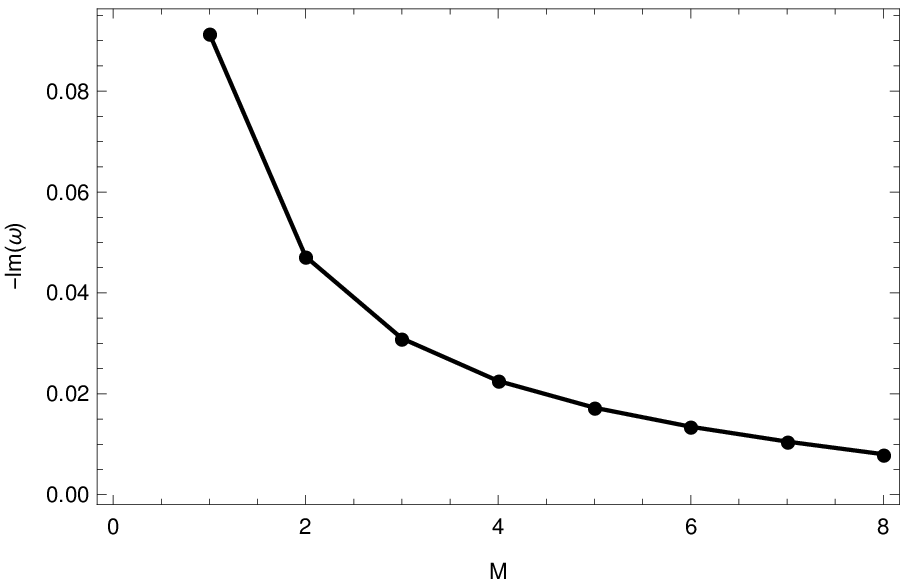}
     \caption{Re $\omega$ and Im $\omega$ vs black hole mass $M$}
     \label{fig2}
     \end{figure}

In the above expression $b=n+\frac{1}{2}$ , $V_0^{(n)}=d^nV/dr_*^n$ at $r=r_0$ and
$\Lambda_4$, $\Lambda_5$ and $\Lambda_6$ can be found in the Appendix of \cite{kon5}. 
The above method also works extremely well in the eikonal limit of large $\ell$ corresponding to large quality
factor, which will also be discussed in the paper. Using Eqn (\ref{qnmeqn}), we computed the QNMs and in \figurename{\ref{fig2}} we plotted
Re $\omega$ and magnitude of Im $\omega$ vs black hole mass. Both Re $\omega$ and Im $\omega$ decreases when
mass $M$ is increased.\par
     \begin{figure*}[htbp]
\centering
\begin{minipage}{.5\textwidth}
  \centering
  \includegraphics[scale=0.66]{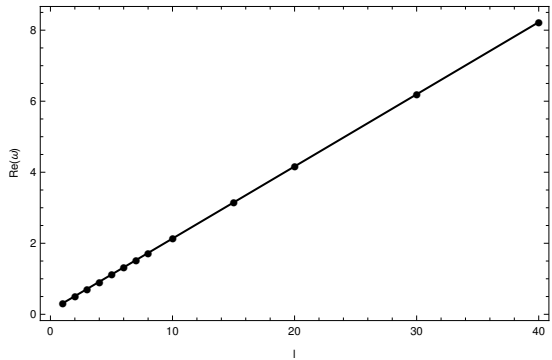}
\end{minipage}%
\begin{minipage}{.5\textwidth}
  \centering
   \includegraphics[scale=0.555]{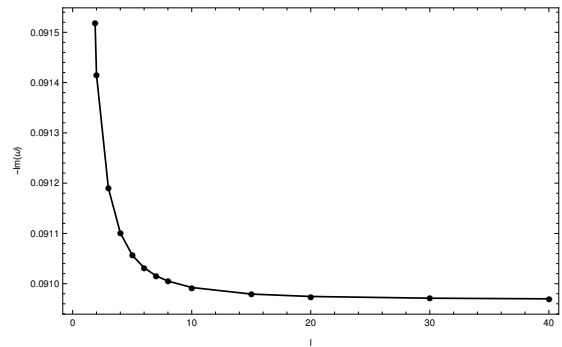}
\end{minipage}
        \caption{Variation of Re $\omega$ and Im $\omega$  with multipole number $\ell$ . Here M=1 , g=0.55, $\Lambda=0.001$.}
        \label{fig3}
\end{figure*}
\begin{figure*}[!htbp]
\centering
\begin{minipage}{.5\textwidth}
  \centering
  \includegraphics[scale=0.85]{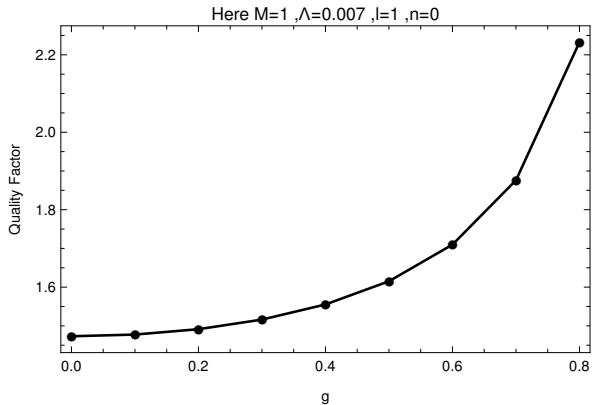}
\end{minipage}%
\begin{minipage}{.5\textwidth}
  \centering
   \includegraphics[scale=0.85]{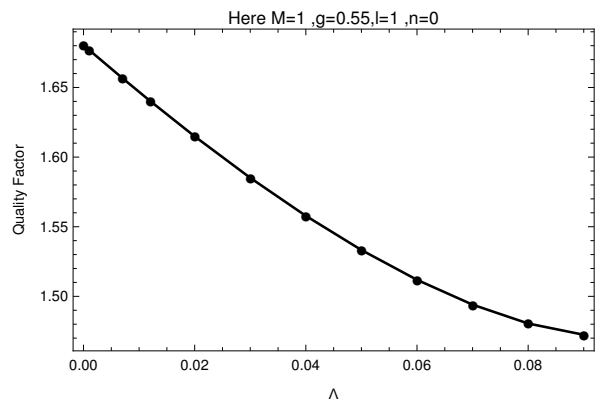}
\end{minipage}
\caption{Q-factor vs parameters $\Lambda$ and $g$.}
\label{qfac}
\end{figure*}
 In TABLE-I, we list the values of the QN frequencies obtained by using third order and sixth order
 WKB approach for the parameter range $\Lambda=0.007$ and $g=0.57$.
 The data from the table suggests that the value of the real part of the frequency shows a 
 steady increase over its third order outcome but on the other hand, the negative imaginary part
 obtained using sixth order WKB method shows a steady decline when compared to the third order result.

\begin{table*}
\begin{center}
\caption{The QN frequencies due to massless scalar perturbation for multipole number $\ell$ ranging between $1$ to $5$ with $\Lambda=0.007$ , and $g=0.57$.}
 \vspace{0.4cm}
 \begin{tabular}{ p{3cm}p{3cm}p{3.5cm}p{3.5cm} }
 \hline
 \hline
  multipole number &Overtone  & 3rd order WKB &6th order WKB\\
 \hline\hline\vspace{0.1cm}\\
 $\ell$=1&n=0 & 0.300446 -0.089967i & 0.302242 -0.090150i  \\
    &n=1 & 0.278912 -0.278097i & 0.282993 -0.277074i\\ \vspace{0.1cm}\\
   \hline \vspace{0.1cm}\\
    &n=0 & 0.499385 -0.088861i & 0.499841 -0.088903i  \\
$\ell$=2&n=1 & 0.485040 -0.269800i & 0.486281 -0.269658i  \\
    &n=2 & 0.461291 -0.456456i & 0.462177 -0.458553i \\ \vspace{0.1cm}\\
    \hline\vspace{0.1cm}\\
    &n=0 & 0.698242 -0.088552i & 0.698417 -0.088563i  \\
$ \ell$=3&n=1 & 0.687778 -0.267316i & 0.688277 -0.267273i  \\
    &n=2 & 0.669085 -0.449812i & 0.669173 -0.450547i  \\
    &n=3 & 0.644523 -0.635942i & 0.643394 -0.640730i \\ \vspace{0.1cm}\\
    \hline\vspace{0.1cm}\\
    &n=0 & 0.897184 -0.088421i & 0.897268 -0.088426i  \\
    &n=1 & 0.888985 -0.266272i & 0.889230 -0.266255i  \\
 $\ell$=4&n=2 & 0.873746 -0.446645i & 0.873717 -0.446952i  \\
    &n=3 & 0.853024 -0.629992i & 0.851862 -0.632179i  \\
    &n=4 & 0.828001 -0.815925i & 0.825285 -0.823178i \\ \vspace{0.1cm} \\
    \hline\vspace{0.1cm}\\
    &n=0 & 1.09619 -0.088354i& 1.09624 -0.088356i  \\
    &n=1 & 1.08946 -0.265738i & 1.08960 -0.265730i \\
    &n=2 & 1.07666 -0.444914i & 1.07663 -0.445061i \\
 $\ell$=5&n=3 & 1.05883 -0.626438i & 1.05796 -0.627545i \\
    &n=4 & 1.03691 -0.810304i & 1.03452 -0.814191i \\
    &n=5 & 1.01159 -0.996158i & 1.00748 -1.005740i \\\vspace{0.1cm}\\
 \hline
\end{tabular}
\end{center}
\end{table*}

\begin{figure*}[htbp]
\centering
\begin{minipage}{.5\textwidth}
  \centering
  \includegraphics[scale=0.58]{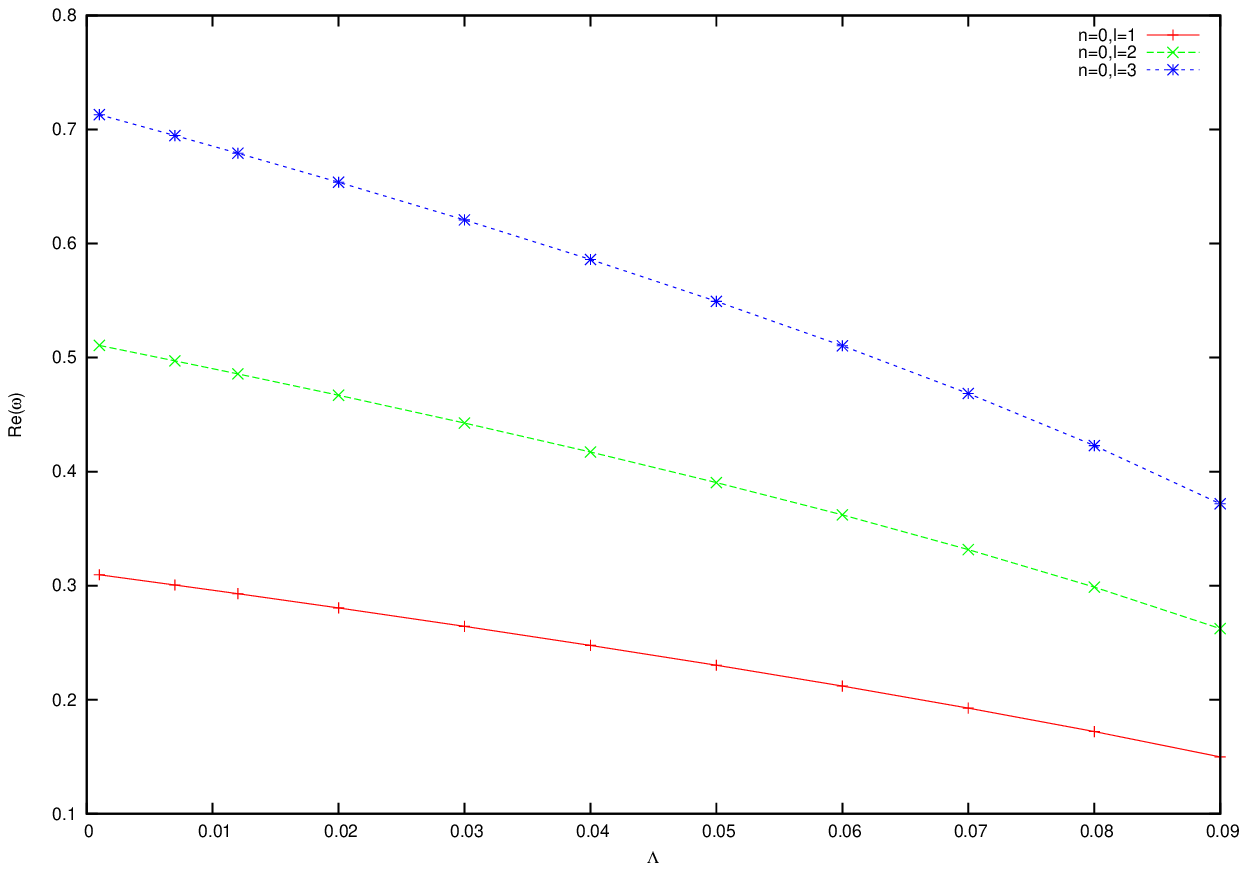}
\end{minipage}%
\begin{minipage}{.5\textwidth}
  \centering
   \includegraphics[scale=0.58]{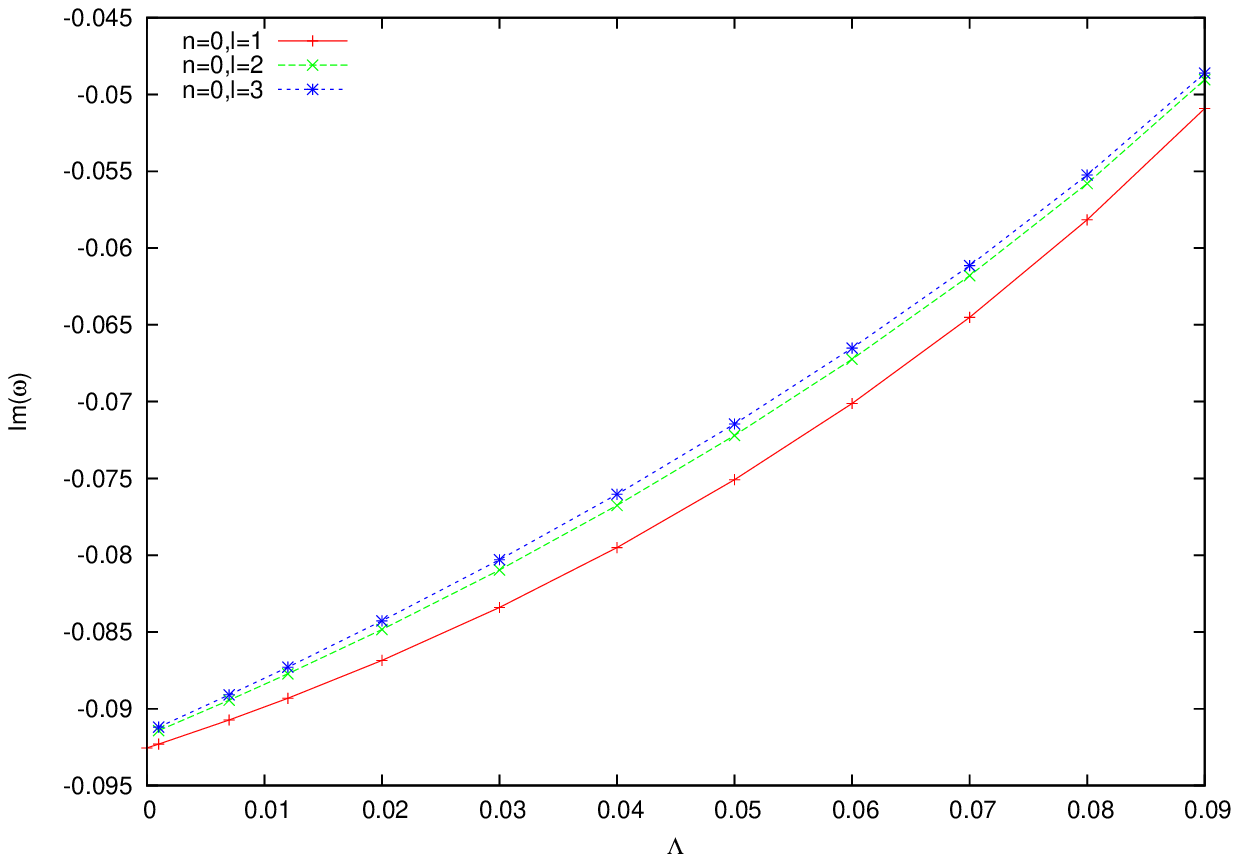}
\end{minipage}
        \caption{ Re $\omega$ and Im $\omega$ vs $\Lambda$ for $g=0.55$ and $M=1$ }
        \label{ReImlambda}
\end{figure*}

\begin{figure*}[htbp]
\centering
\begin{minipage}{.5\textwidth}
  \centering
  \includegraphics[scale=0.58]{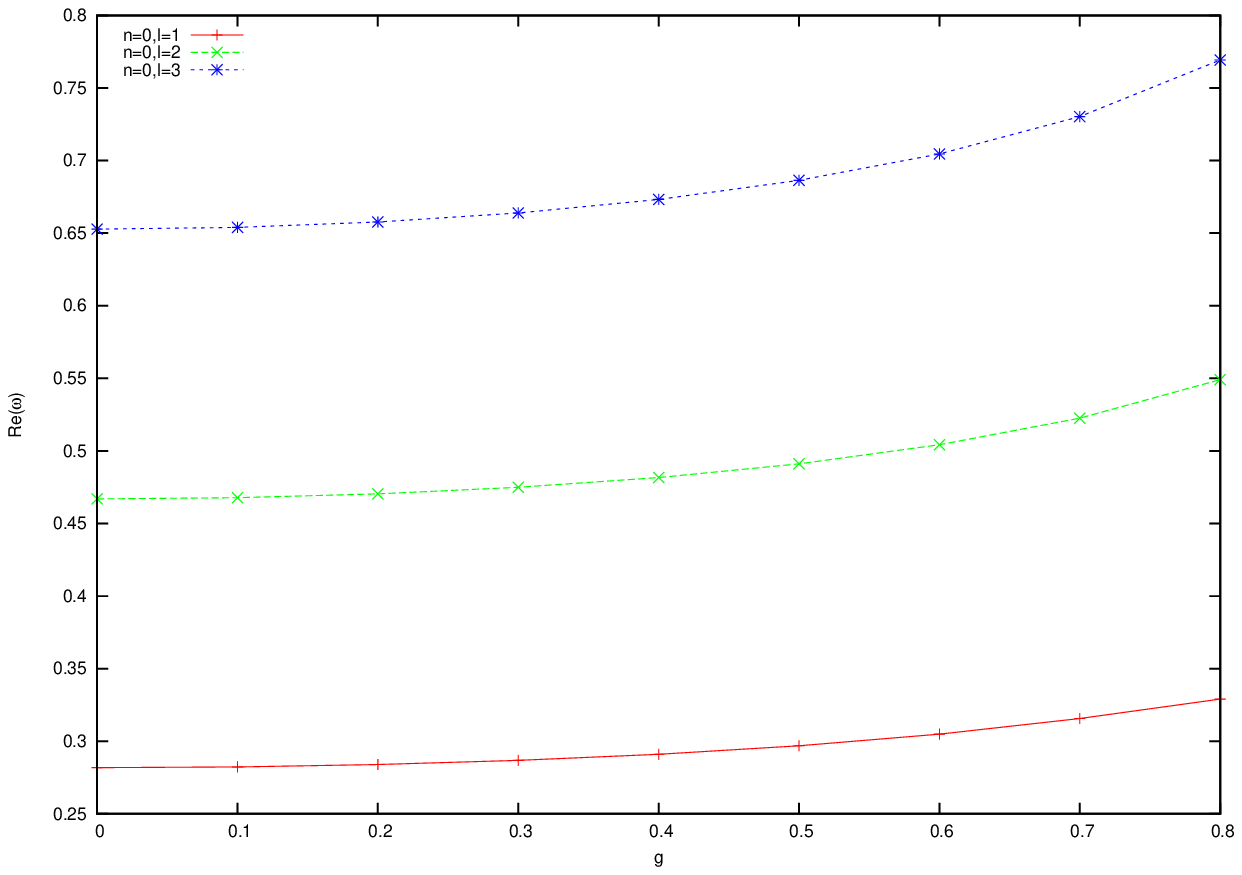}
\end{minipage}%
\begin{minipage}{.5\textwidth}
  \centering
   \includegraphics[scale=0.58]{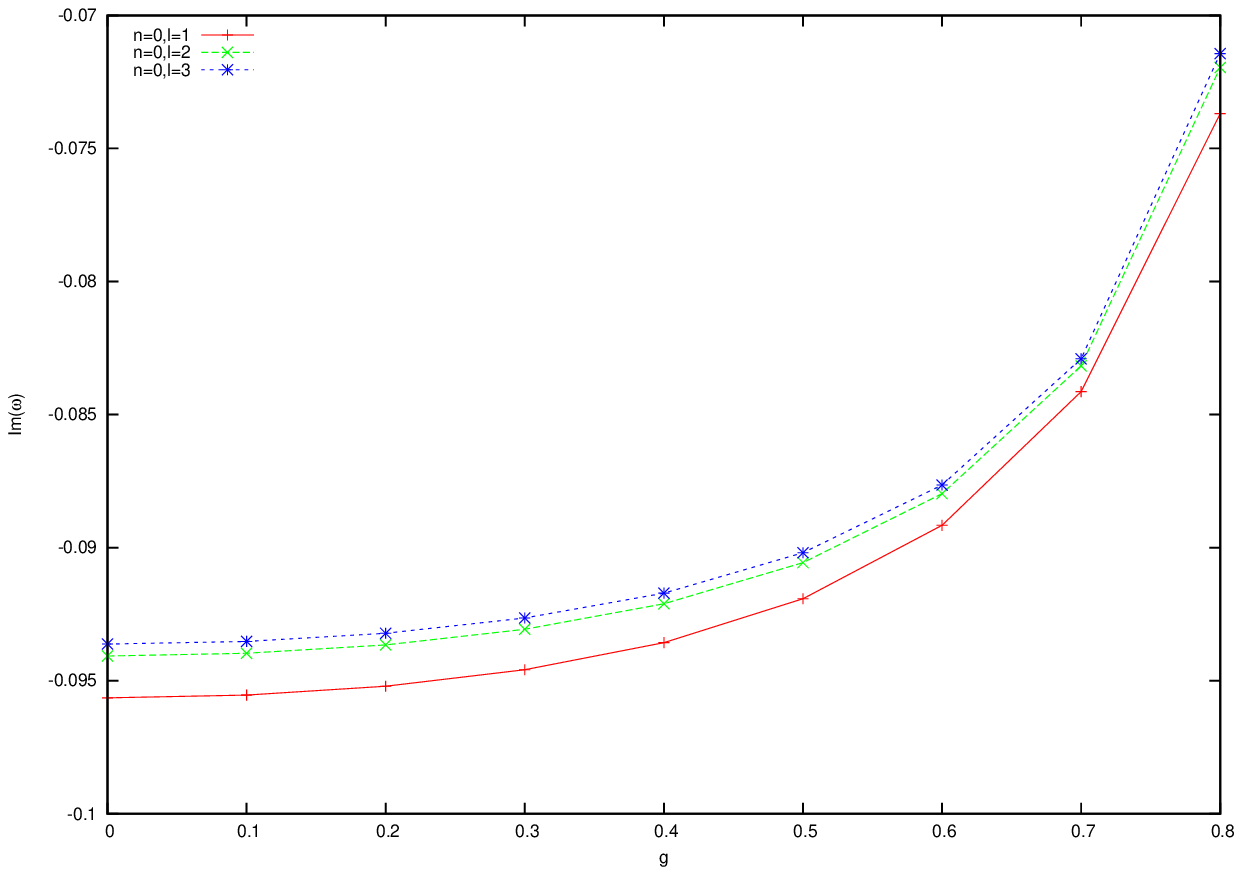}
\end{minipage}
       \caption{Re $\omega$ and Im $\omega$ vs $g$ for $\Lambda=0.007$ and $M=1$}
       \label{ReImg}
       \end{figure*}

In \figurename{\ref{ReImlambda}} \& \figurename{\ref{ReImg}} we plot the behaviour of low lying QN frequencies vs $\Lambda$ and $g$ for different $\ell$ .
Both the plots reveals that Re $\omega$ and Im $\omega$ decreases with increasing $\Lambda$. Real part of frequencies
still increasing steadily with g increased and Imaginary part decreases in magnitude. We have also
computed the QN frequencies for larger multipole number $\ell$ with  overtone $n=0$ only.\par
We plot for $\ell$ ranging between 1 to 40 while we have 
  fixed the values of $\Lambda=0.001$, $g=0.55$ and $n=0$. Re($\omega$) increases linearly with $\ell$\cite{qnmbbh} while magnitude of
  Im($\omega$) first decreases and remains constant for larger $\ell$.
  
To examine the field oscillations, we will define the Quality Factor(Q.F) as
\begin{align}
Q.F = \frac{Re(\omega)}{2|Im(\omega)|}
\end{align}
We plotted the Q.F versus the parameters $\Lambda$ and $g$ in \figurename{\ref{qfac}}.
Quality Factor increases with increasing $g$ and decreases with an increase in $\Lambda$. Thus, Q.F implies that oscillations will be 
more with larger magnetic charge $g$ and decay faster for small $\Lambda$.\par

It is worth mentioning here that by computing the Lyapunov exponent (the inverse
of the instability timescale associated with the geodesic motion), one can show that, in the eikonal
limit, QNMs of black holes in any dimensions are determined by the parameters of the circular null geodesics \cite{cbwz}. This is a very strong result and is independent of the field equations. The only assumption goes into the calculation is the fact that the black hole spacetime is static, spherically symmetric and asymptotically flat. However a non-trivial example of non-asymptotically flat near extremal Schwarzschild de Sitter black hole space time was also discussed in this context. The same argument can be applied in case of BdS black holes too in the limit of near extremal Nariai or cold black holes where either the black hole horizon and the cosmological horizon merges or the inner and outer horizon coincides. In these limits it may be possible to get the eikonal limit using the WKB method following \cite{cbwz}. 

\subsection{Massive scalar field perturbations}
It is by now clear that perturbations of black holes due to presence of a scalar field in the background are mainly of theoretical interest. However, these kind of perturbations could be observationally relevant if boson stars can be proven to be a possible dark matter source \cite{sw}. 
Boson stars are objects made up of self-gravitating scalar fields. If they become unstable and collapse to form a black hole, one should expect both gravitational and scalar waves to be emitted with appropriate quasi-normal modes. This in general means that the study of massive scalar quasinormal modes is also another important issue which may provide a toy model for the study of a bigger picture. On the other hand, the behaviour of the massive scalar field in a black hole background is completely different from that of a massless one: firstly, massive scalars demonstrates the so called superradiant instability and secondly, at asymptotically late times, the massive fields show universal behaviour irrespective of the spin of the field.

\begin{figure}[htbp]
\centering
 \includegraphics[width=7.2cm]{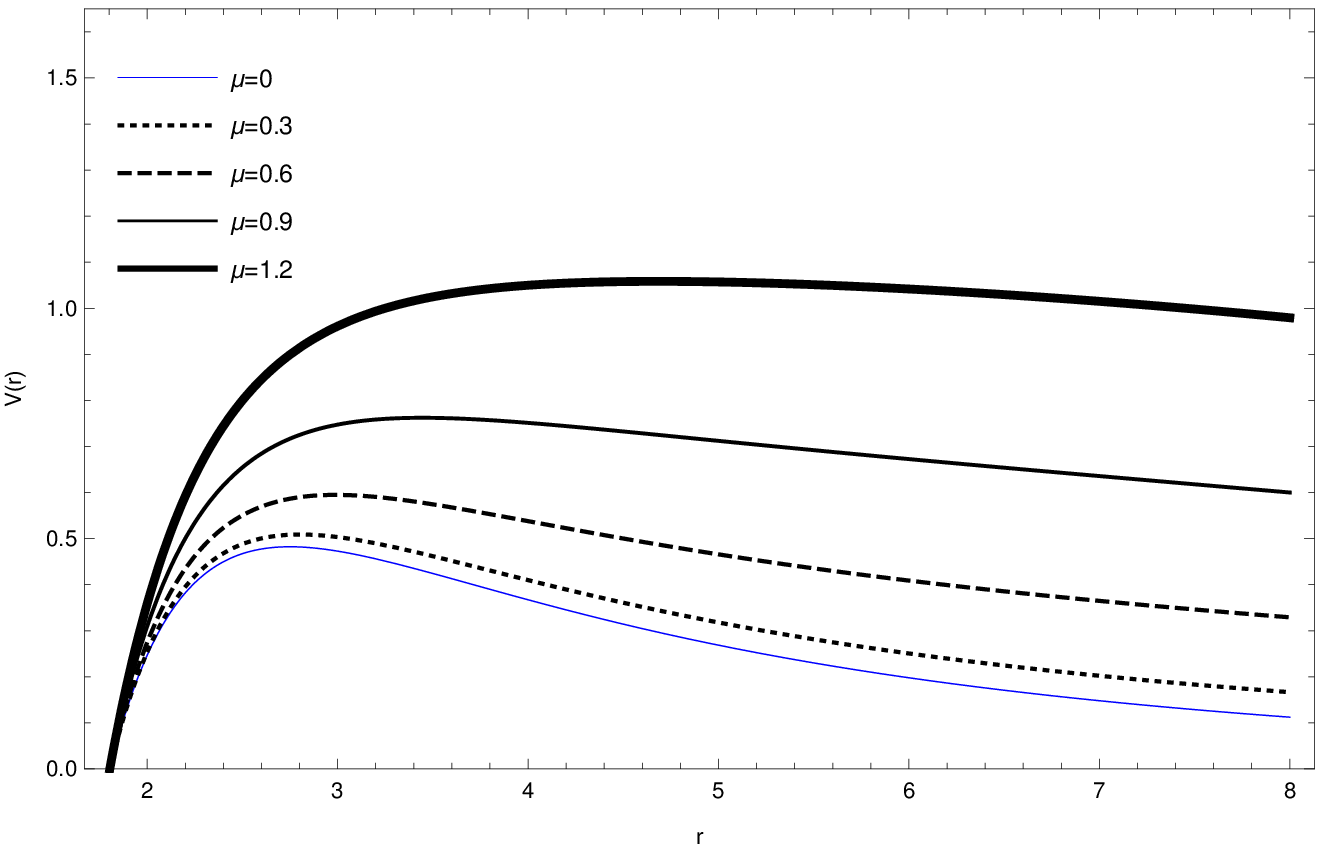}
\caption{Variation of Potential V(r) vs $r$ for various masses($\mu$)}
\label{potms}
\end{figure}

For massive scalar perturbation the Klein Gordon equation is given by
\begin{align}
  \frac{1}{\sqrt{-g}} \partial_{\mu}\left(\sqrt{-g}g^{\mu\nu}\partial_\nu \Phi\right)= \mu^2\Phi
\end{align}
 where $\mu$ is scalar field mass. Similarly, we chosen the ansatz as in equation (10) and finally we have the Schr\"{o}dinger-like equation and modified
 effective potential as
 \begin{align}
  &\frac{d^2U(r)}{dr_*^2}+\left(\omega^2-V(r_*)\right)U(r)=0\nonumber\\
  &V(r)=f(r)\left(\frac{l(l+1)}{r^2}+\frac{f'(r)}{r}+\mu^2\right)
 \end{align}
where the tortoise coordinate $r_*$ is related to $r$ by $dr_*=\frac{dr}{f(r)}$.
 In \figurename{\ref{potms}}, we plot the effective potential $V(r)$ vs $r$ for different scalar mass($\mu$). We have chosen the
parameters $\Lambda = 0.007$ , $M = 1$ , $g = 0.5$ and $\ell = 3.$

Notice that the peak of the potential depends on the
scalar field mass $\mu$ with other parameters fixed. Since QNMs are known to be the waves trapped within the
peak of this potential \cite{will}. As discussed in \cite{aki}, we expect
similar behaviour for BdS black hole that the imaginary
part of the quasinormal modes frequencies will decrease
for large $\mu$. However, the real part of QNMs will increase
as $\mu$ increases.\par

\begin{figure*}[htbp]
\centering
\begin{minipage}{.5\textwidth}
  \centering
 \includegraphics[width=7.5cm]{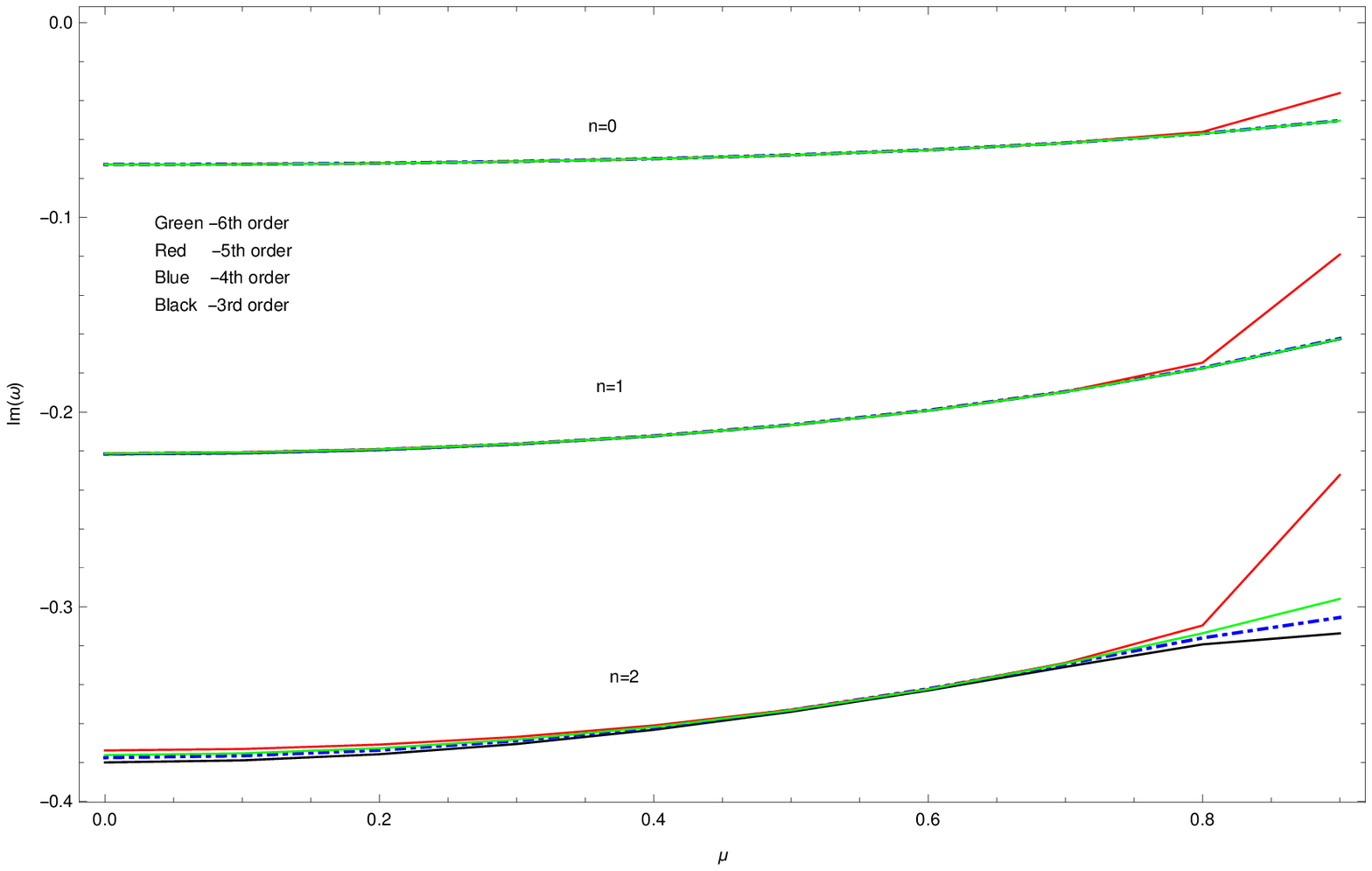}\vspace{0.2cm}
\end{minipage}%
\begin{minipage}{.5\textwidth}
  \centering
 \includegraphics[width=7.5cm]{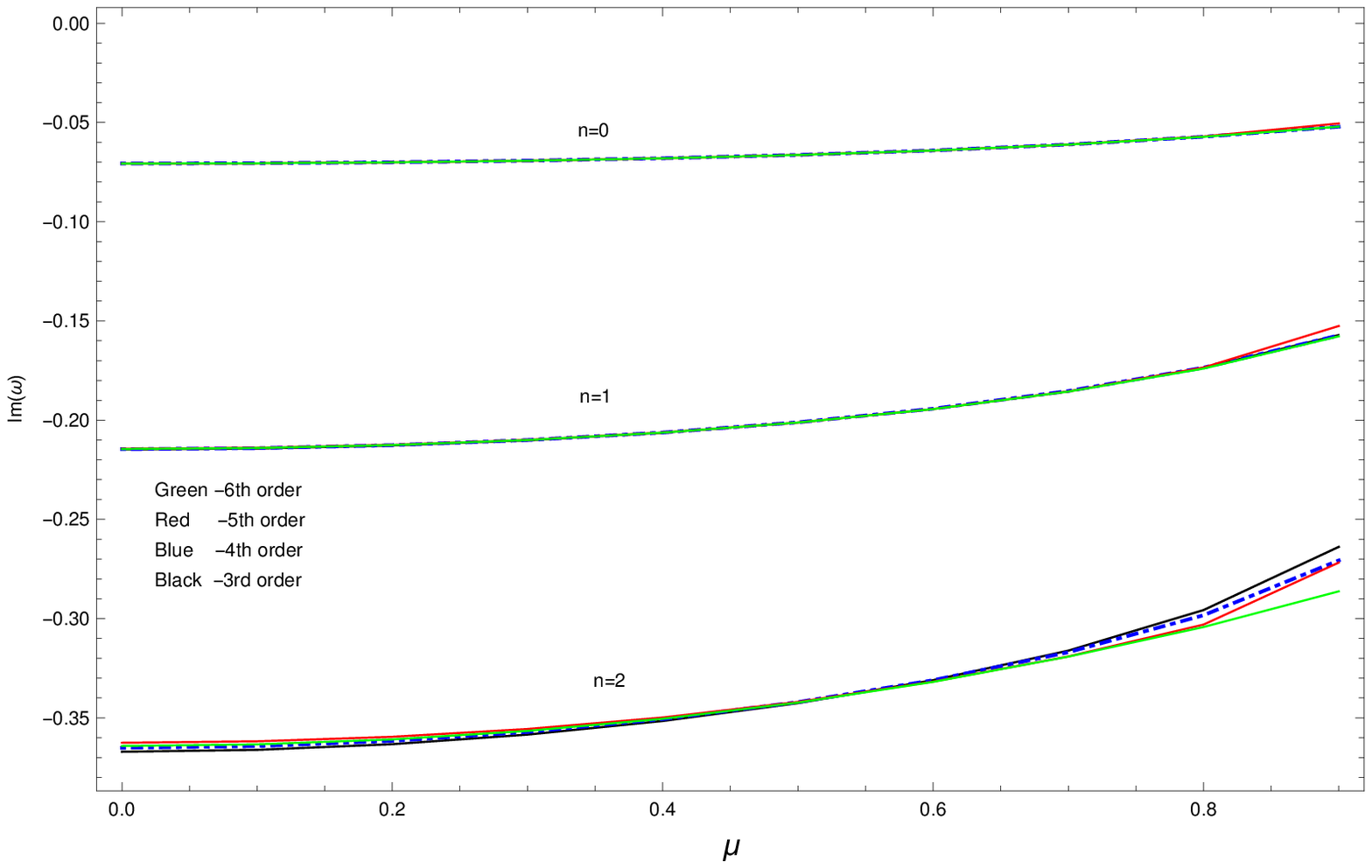}\vspace{0.2cm}
\end{minipage}\\
\caption{Variation of Im($\omega$) with scalar mass$(\mu)$ for $\Lambda=0.001$(left) and $\Lambda=0.01$(right).
Here magnetic charge $g=0.8$ and multipole number $\ell=3$.}
\label{Immu}
\end{figure*}

\begin{figure*}[!htbp]
 \centering
\begin{minipage}{.5\textwidth}
  \centering
 \includegraphics[width=7.5cm]{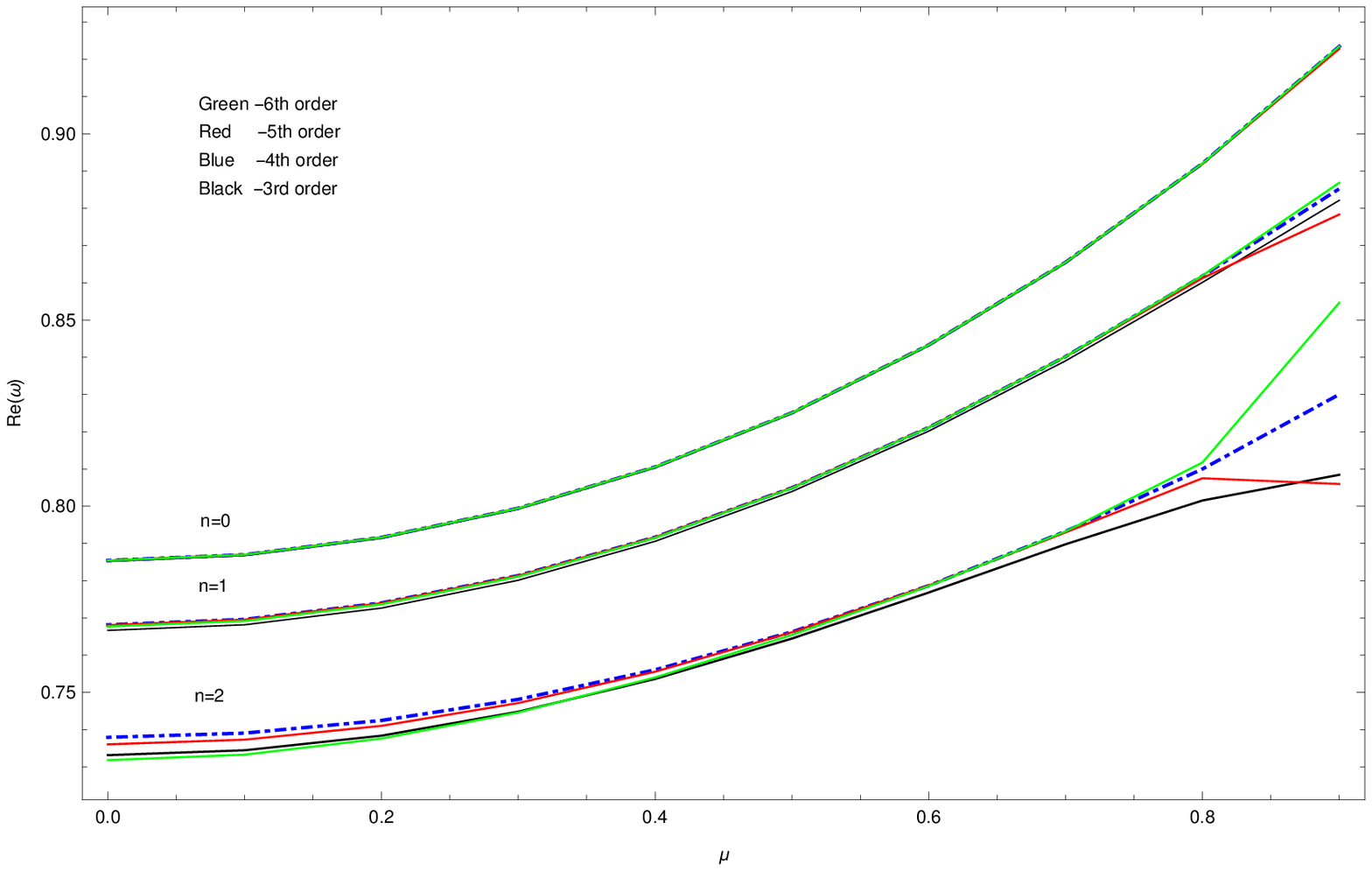}\vspace{0.2cm}
\end{minipage}%
\begin{minipage}{.5\textwidth}
  \centering
 \includegraphics[width=7.5cm]{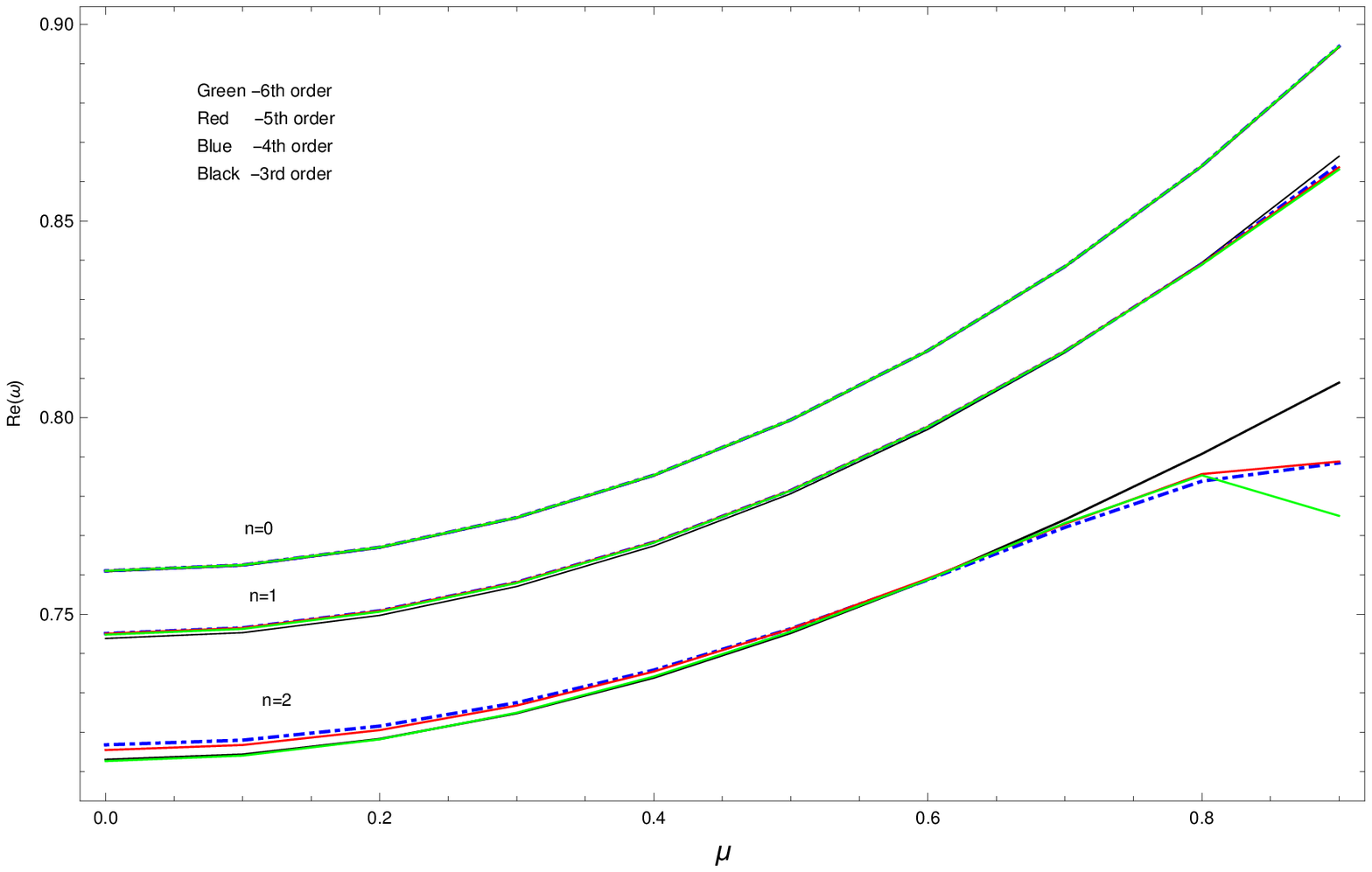}\vspace{0.2cm}
\end{minipage}
\caption{Variation of Im($\omega$) with scalar mass$(\mu)$ for $\Lambda=0.001$(left) and $\Lambda=0.01$(right).
Here magnetic charge $g=0.8$ and multipole number $\ell=3$.}
\label{Remu}
\end{figure*}

In \figurename{\ref{Immu}} \& \figurename{\ref{Remu}}, we have plotted the variations of imaginary and real part of $\omega$ versus scalar field mass $(\mu)$ for different values
of parameters $\Lambda$ , $M$ , $g$ and $l$. We have plotted all the data obtained by 3rd, 4th, 5th and 6th  
order WKB calculations simultaneously to compare the accuracy between different orders. We observed from the plots of both Im($\omega$) and Re($\omega$) that for low overtone number $n$,
the accuracy between lower and higher order WKB is not much significant but for large $n$ deviation is more. The magnitude of Im($\omega$) decreases with increasing scalar mass, on the contrary, the magnitude of Re($\omega$) increases with increasing field mass. \par
In TABLE-II, we present the numerical values of QN frequencies with corresponding parameters. Since it is well known that
WKB method is more accurate for $n<\ell$ , we have tabulated the QNMs frequencies for $n<\ell$ only.

\begin{table*}[htbp]
  \begin{center}
  \caption{ Comparison between 3rd and 6th order WKB of QN frequencies due to massive scalar perturbation for $\Lambda=0.001$ , $g=0.8$ , $M=1$ and multipole number $\ell = 3 .$}

\vspace{0.4cm}
\begin{tabular}{ p{3cm}p{3cm}p{3.5cm}p{3.5cm} }
 \hline
  scalar mass & Overtone number & 3rd order WKB &6th order WKB\\
 \hline
 $\mu$=0&$n=0$ & 0.785191 -0.072860 i &  0.785427 -0.072871 i \\
 &$n=1$ & 0.766662 -0.221907 i &  0.767696 -0.221308 i \\
 &$n=2$ & 0.733146 -0.379935 i & 0.731816 -0.376151 i  \vspace{0.2cm}\\
 \hline
 $\mu$=0.2&$n=0$ & 0.791386 -0.072173 i & 0.791621 -0.072187 i  \\
 &$n=1$ & 0.772648 -0.219660 i & 0.773657 -0.219163 i  \\
 &$n=2$ & 0.738382 -0.375726 i &  0.737619 -0.372627 i \vspace{0.2cm}\\
 \hline
 $\mu$=0.4&$n=0$ & 0.810314 -0.069906 i & 0.810545 -0.069926 i  \\
 &$n=1$ & 0.790571 -0.212544 i & 0.791508 -0.212286 i  \\
 &$n=2$ & 0.753616 -0.363179 i & 0.754017 -0.361725 i  \vspace{0.2cm}\\
 \hline
 $\mu$=0.6&$n=0$ & 0.843078 -0.065359 i & 0.843303 -0.065384 i  \\
 &$n=1$ & 0.820174 -0.199327 i & 0.821114 -0.199253 i  \\
 &$n=2$ & 0.776829 -0.342952 i & 0.778539 -0.342391 i \vspace{0.2cm} \\
 \hline
 $\mu$=0.8& $n=0$ & 0.891821 -0.056961 i & 0.892032 -0.057001 i  \\
 &$n=1$& 0.860107 -0.177635 i & 0.861930 -0.177404 i \\
 &$n=2$& 0.801532 -0.319314 i & 0.811731 -0.313581 i \\
 \hline
\end{tabular}
\end{center}
\end{table*}\vspace{0.2cm}
 
\begin{figure}[htbp]
\centering
 \includegraphics[width=6.6cm]{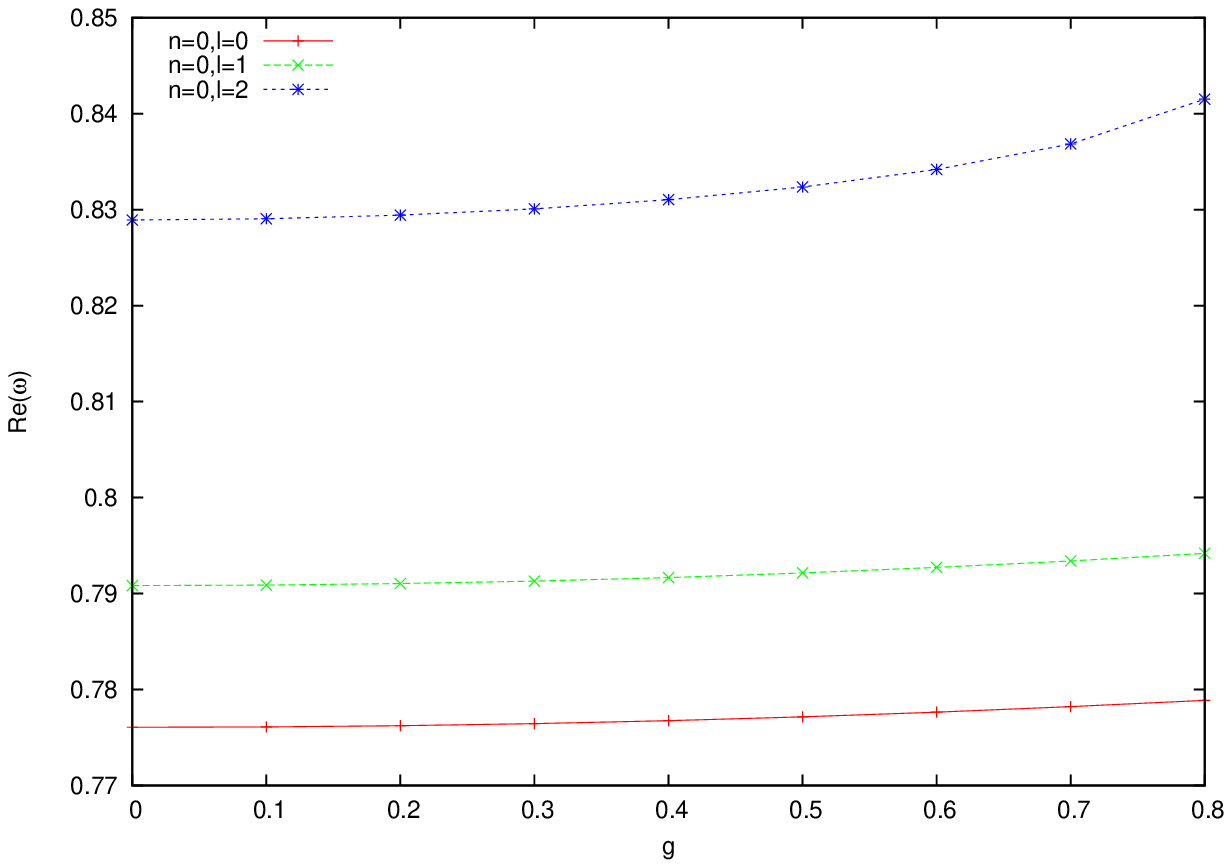}\\
 \includegraphics[width=6.6cm]{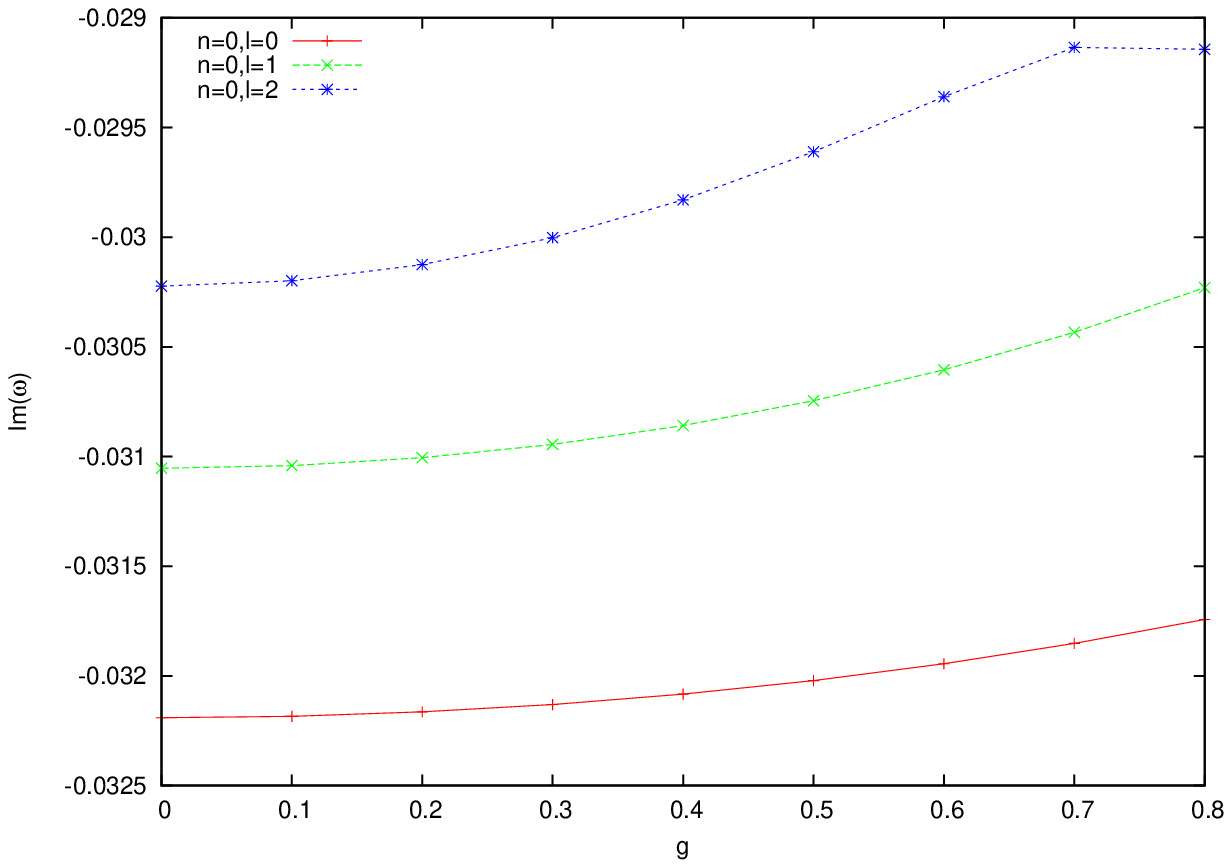}
 \caption{ Variation of Re($\omega$) and Im($\omega$) with $g$, for $M=1$, $\Lambda=0.007$ .}
 \label{figure10}
\end{figure}

\begin{figure}[htbp]
\centering
 \includegraphics[scale=0.52]{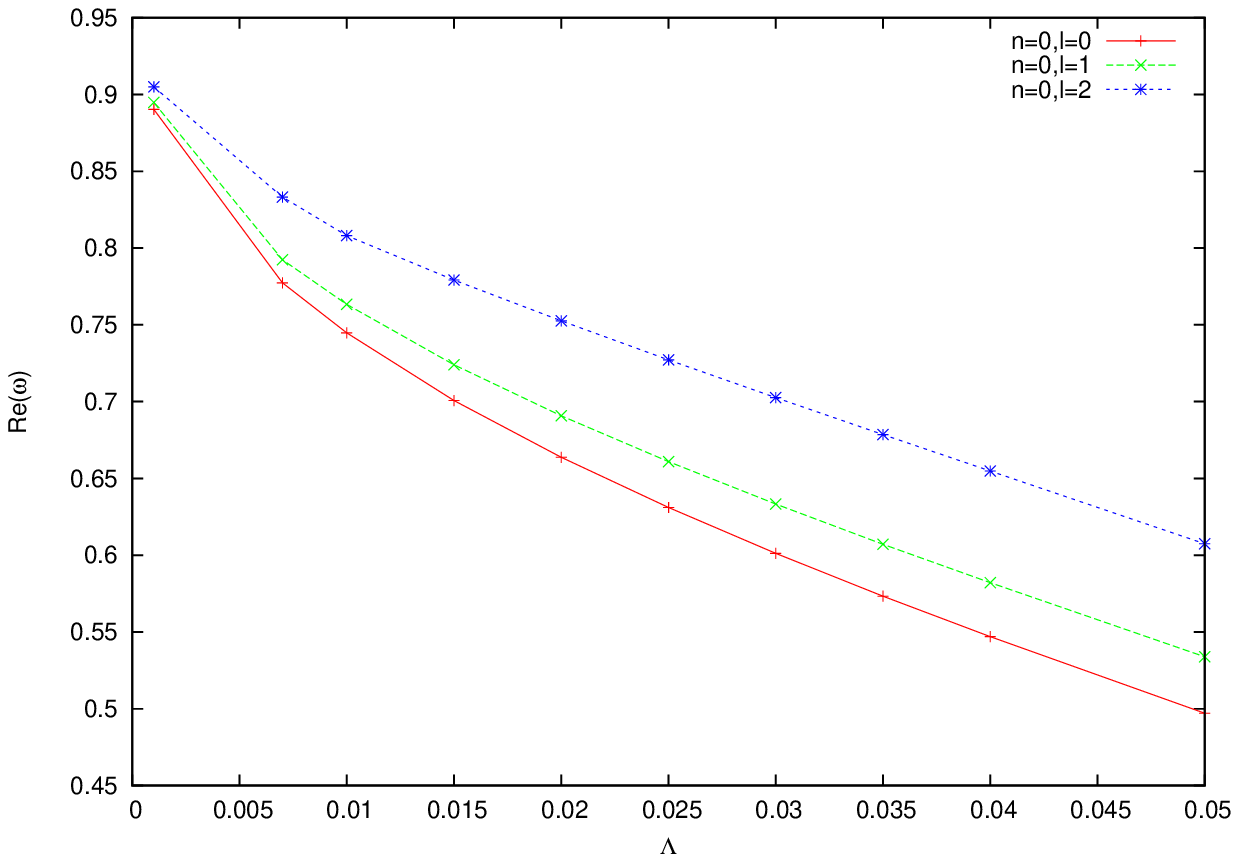}\\
 \includegraphics[scale=0.52]{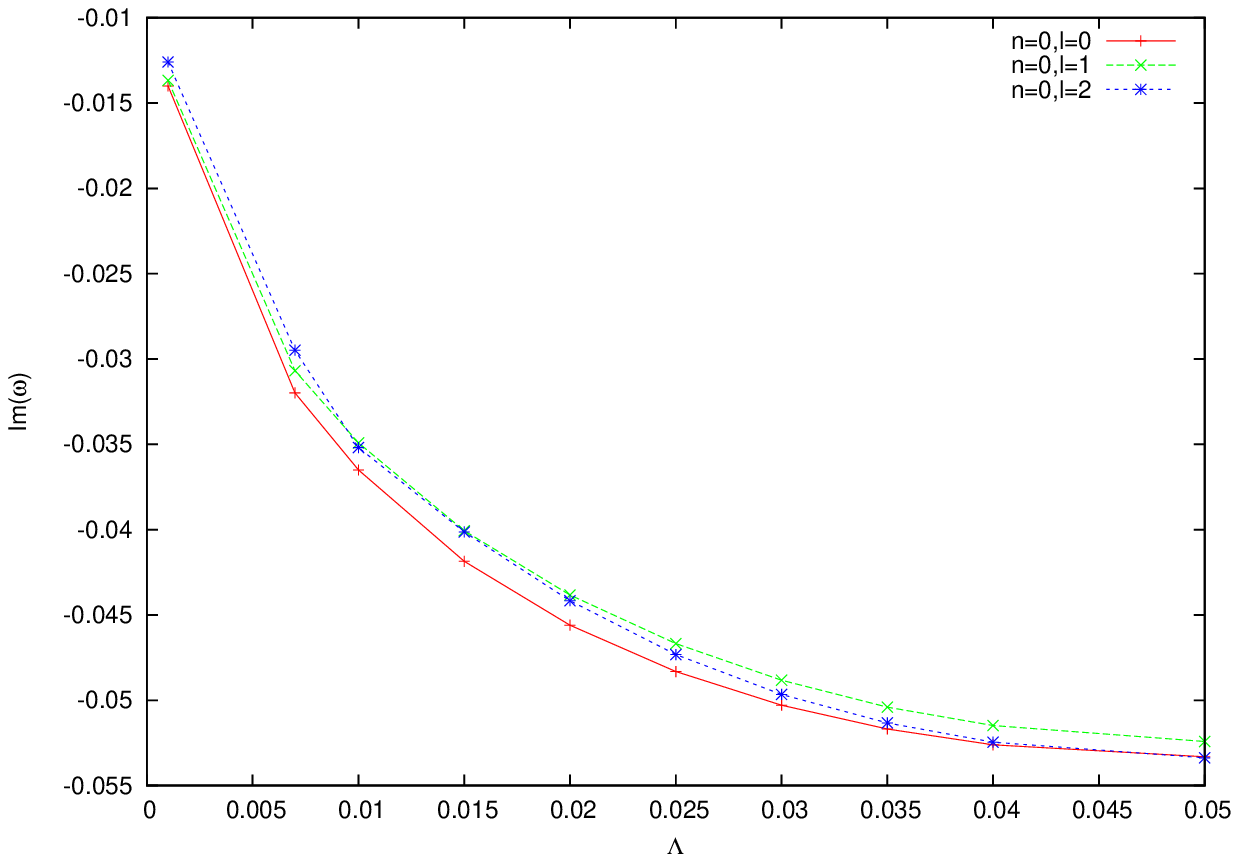}
 \caption{ Variation of Re($\omega$) and Im($\omega$) with $\Lambda$, for $M=1$, $g=0.55$ .}
 \label{figure11}
\end{figure}

As in massless case, we plot in \figurename{\ref{figure10}} \& \figurename{\ref{figure11}} the behaviour of QN frequencies with
$\Lambda$ and $g$ for $n=0$ and $\ell=0,1,2$ respectively. Re($\omega$) increases with increasing value of magnetic charge $g$ while magnitude Im($\omega$) decreases.
This behaviour of QNMs can be well understood from the form of the potential. As the height of the potential peak increases with $g$, therefore real part of QNMs increases.\par
 On the other hand,\figurename{\ref{figure11}} shows that Re($\omega$) decreases with an increase in cosmological constant($\Lambda$) but
Im($\omega$) increases with $\Lambda$ in magnitude. Similarly if we plot the variation of potential with $\Lambda$ the height decreases, thus Im ($\omega$) increases.
Hence, we can say for scalar field perturbations with the scalar mass included, the oscillations decay faster with large cosmological constant
$\Lambda$ and
oscillates better for large magnetic charge $g$. \par

\section{Dirac QNMs in BdS black hole}

In this section, we will extend our discussion to massless Dirac perturbations for BdS black holes. 
As in \cite{Brill:1957fx}, by starting from Dirac equation in spherically symmetric curved background,
the Schr\"{o}dinger-like  equation we finally arrived at is given by
\begin{align}
\left(-\frac{d}{dr^2_*}+V_1\right)G&=E^2G \label{sc1}\hspace{0.2cm}\\ \hspace{0.2cm} \left(-\frac{d}{dr^2_*}+V_2\right)F&=E^2F \label{sc2}
\end{align}
where $r_*$ is the tortoise coordinate given by $f\frac{d}{dr}\equiv\frac{d}{dr_*}$, $E$ is the energy. The effective potentials is given by
\begin{align}
 V_{1,2}=\pm\frac{dW}{dr_*}+W^2\hspace{0.2cm} , {\rm{where}} \hspace{0.3cm}W=\frac{\sqrt{f}}{r} (\ell+1)
\end{align}

A detailed discussion about the derivation of the above equations can be found in the appendix of \cite{skc11}. It is worth mentioning here that the potentials $V_1$ and $V_2$ corresponding to Dirac particles and anti-particles are supersymmetric to each other and derived from the same
superpotential $W$.We will evaluate the quasinormal modes by solving equation (\ref{sc1}) taking only $V_1$ as it is well known that both Dirac particles and anti-particles have the same quasi-normal spectra \cite{jing}. Therefore, in the context of perturbation, it does not matter if 
one perturbs the black hole with a particle or an anti-particle because of this isospectrality of QN spectra. \par
\begin{figure}[htbp]
 \includegraphics[width=8cm]{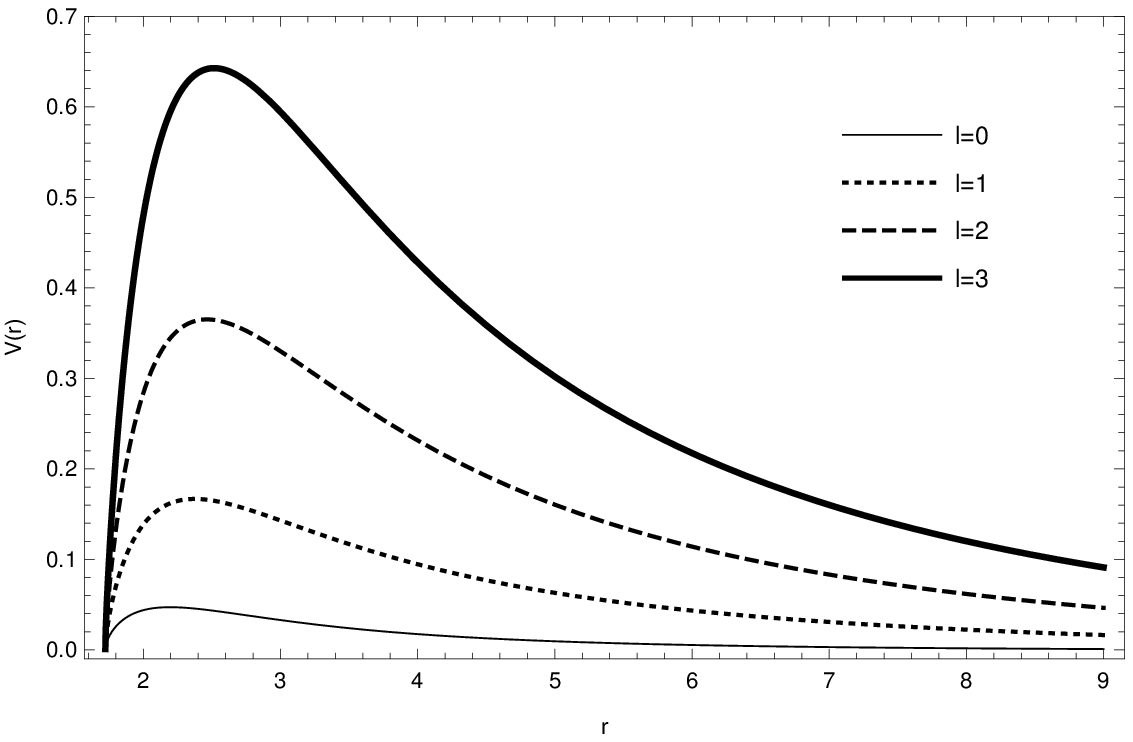}
\caption{ Variation of V($r$) vs $r$ for different values of $\ell$ for the massless Dirac perturbations.}
\label{fig12}
\end{figure}

In \figurename{\ref{fig12}}, we showed the behaviour of the effective potential ($V_1$ only) for BdS black hole with spherical harmonic $l$ 
for parameters $ M=1, \Lambda=0.007, g=0.57.$
 
We have computed the massless fermion QNMs semi-analytically using sixth order WKB method. The plots are shown below. In \figurename{\ref{fig13}}, we showed the variation of real
and imaginary part of $\omega$ with cosmological constant $\Lambda$ and in \figurename{\ref{fig14}}, the variation
with magnetic charge $g$ for different values of $\ell$ with fixed overtone number($n=0$) are
shown. We can clearly see from the plots that Re($\omega$) slowly increases with an increase in the magnetic charge $g$ of the BdS black hole whereas it slowly decreases with increasing value of $\Lambda$. Whereas the behaviour of the imaginary part of the frequency reverses
its role, i.e. as we increase the cosmological constant, the imaginary part increases, however it decreases if we increase the magnetic charge, keeping all other parameters fixed. 

\begin{figure*}[htbp]
  \centering
  \begin{minipage}{.5\textwidth}
  \centering
 \includegraphics[width=7cm]{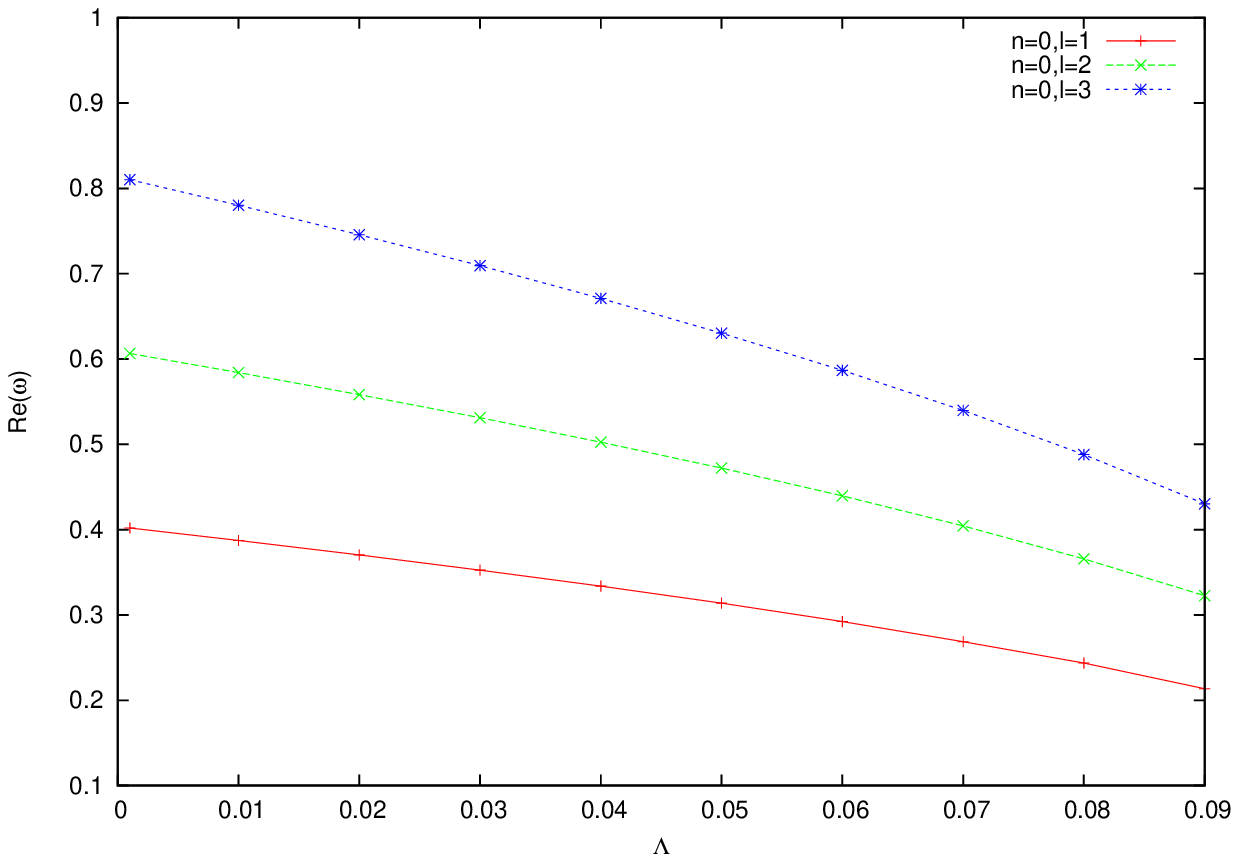}
\end{minipage}%
  \begin{minipage}{.5\textwidth}
  \centering
 \includegraphics[width=7cm]{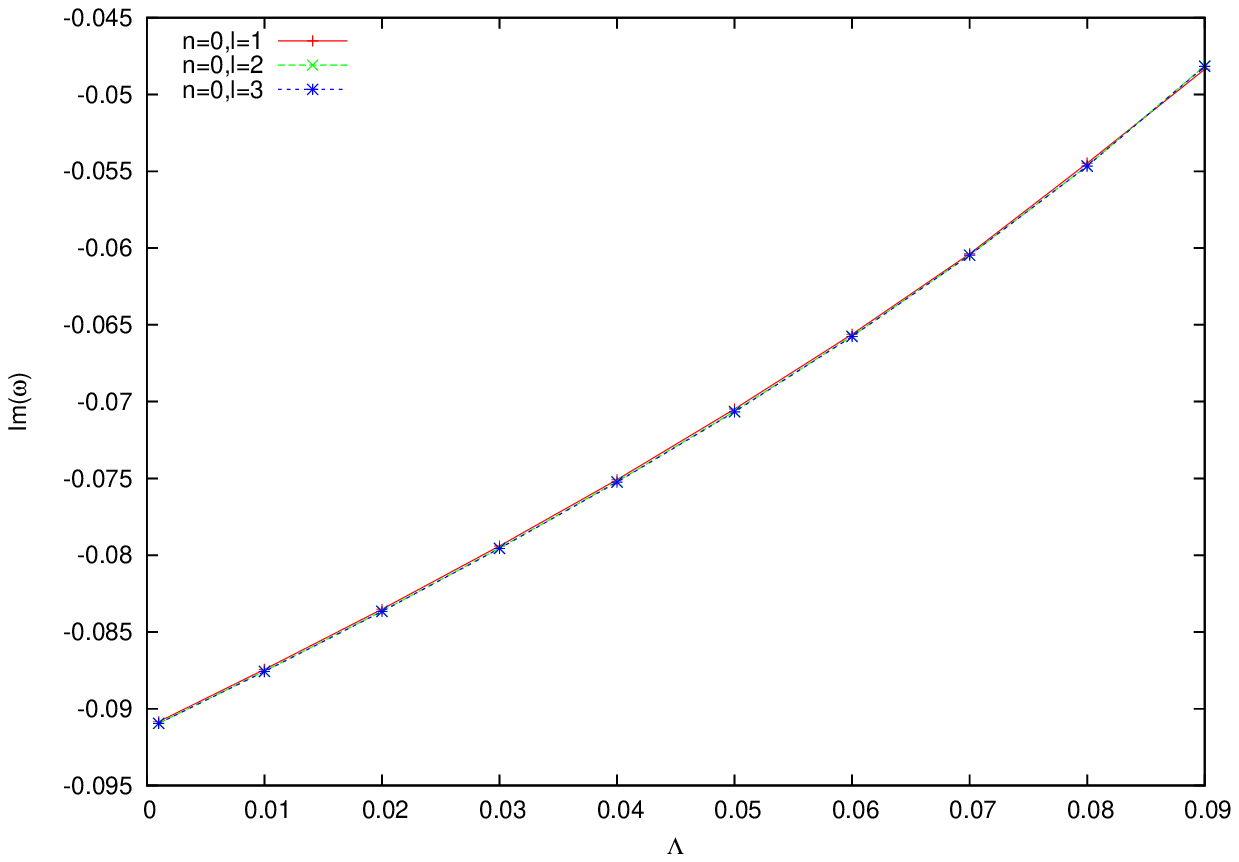}
\end{minipage}
 \caption{Variation of Re($\omega$) and Im($\omega$) vs $\Lambda$, for $M=1$, $g=0.55$ .}
 \label{fig13}
 \end{figure*}

\begin{figure*}[htbp]
  \centering
  \begin{minipage}{.5\textwidth}
  \centering
 \includegraphics[width=7cm]{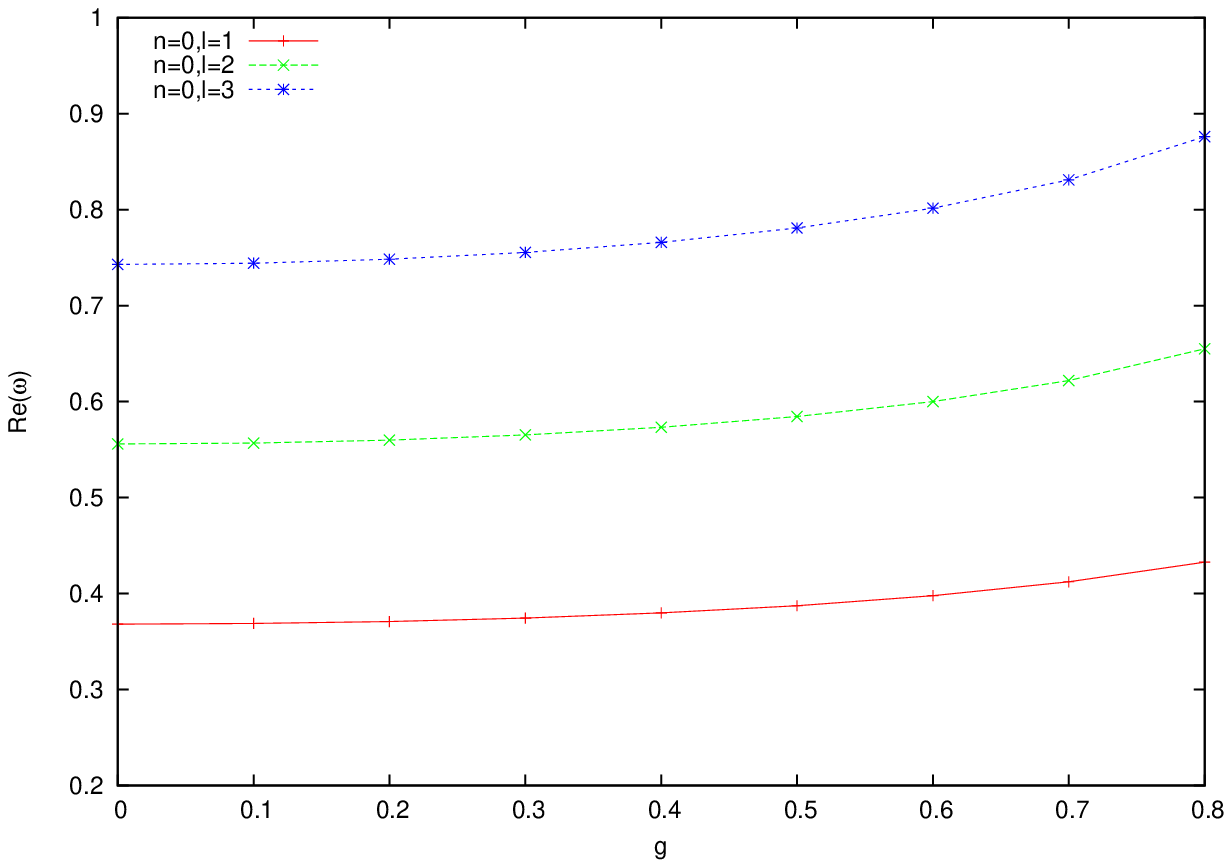}
\end{minipage}%
  \begin{minipage}{.5\textwidth}
  \centering
 \includegraphics[width=7.4cm]{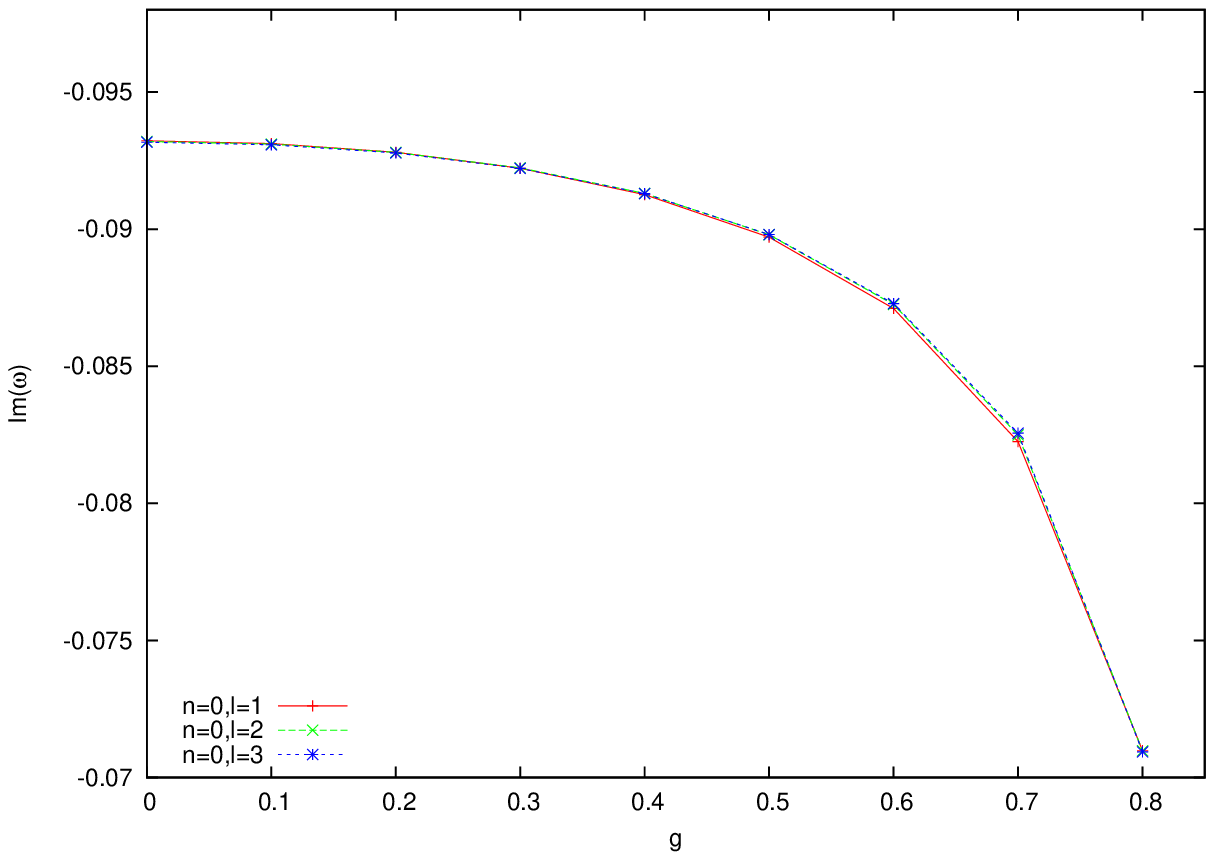}
\end{minipage}
\caption{Variation of Re($\omega$) and Im($\omega$) vs $g$, for $M=1$, $\Lambda=0.007$ .}
\label{fig14}
 \end{figure*}

 
 \section{Summary and Conclusion}
In this paper, we have discussed the  massless and massive scalar field perturbations and 
the massless Fermionic perturbations for a regular BdS black hole. We have used sixth order WKB approximation method to 
calculate the QNMs frequencies. We studied how the frequencies vary  as a function of the scalar field mass ($\mu$), multipole 
number ($\ell$) as well as with the parameters like the cosmological constant ($\Lambda$), black hole mass ($M$) and magnetic charge  ($g$). We found that the QN frequencies decrease with
an increase in black hole mass \cite{qnmbbh}. The plots of frequencies versus the scalar mass show that Re $\omega$ increases with mass $\mu$
while Im $\omega$ decreases. The figures also suggested that if we plot the frequencies from low to higher overtones taking into account different WKB orders, we see that comparative accuracy is better for $\ell< n$. We also found that Re $\omega$ decreases
with an increase in cosmological constant $\Lambda$ for scalar (both massless and massive) perturbations as well as with Dirac
perturbations but Im $\omega$ decreases in massless and fermionic case however, increases for the massive case when $\Lambda$ 
is increased. We have also studied the behaviour of how the Q-factor for the massless scalar field varies with $\Lambda$ and $g$.\par
For massive scalar perturbations, we see that mass $\mu$ enhances the field oscillations and decreases the damping for small $\Lambda$, 
unlike in the massless case where it is just the opposite. In all the three  scenarios real frequency of oscillations Re $\omega$
increases steadily with magnetic charge ($g$) but the damping denoted by Im $\omega$  decreases.\par
For future directions, it would be interesting to study the time evolution of perturbations for this particular black hole. A look into the gravitational (metric) perturbation of the BdS geometry will also be an interesting topic.  Apart from that, in \cite{split}, the authors have used the conformal properties of the spinor field to obtain the Dirac QNMs for a higher dimensional Schwarzschild-Tangherlini black hole. They have described these modes in the light of the so-called split fermion models which have massive  fermions in the bulk and it will be interesting to study such massive Dirac perturbations in the context of higher-dimensional generalization of the BdS black holes. 

\bibliographystyle{unsrt}


\end{document}